\documentclass[sigconf]{acmart}

\usepackage{array}
\usepackage{xspace}

\definecolor{highlightcolor}{HTML}{000000} 
\newcommand{\highlight}[1]{\textcolor{highlightcolor}{#1}}

\usepackage{listings}
\usepackage{xcolor}
\usepackage{acmart-taps}

\aptLtoXcmd{\newcommand{\inlinecode}[1]{\colorbox{yellow!30}{\textsf{#1}}}}{\newcommand{\inlinecode}[1]{\colorbox{yellow!30}{\lstinline[basicstyle=\ttfamily]|#1|}}}

\newcommand{\incremental}{\textsc{Stepwise}\xspace}
\newcommand{\structural}{\textsc{Phasewise}\xspace}
\newcommand{\conversational}{\textsc{Conversational}\xspace}

\AtBeginDocument{%
  \providecommand\BibTeX{{%
    \normalfont B\kern-0.5em{\scshape i\kern-0.25em b}\kern-0.8em\TeX}}}

\copyrightyear{2024}
\acmYear{2024}
\setcopyright{acmlicensed}\acmConference[UIST '24]{The 37th Annual ACM Symposium on User Interface Software and Technology}{October 13--16, 2024}{Pittsburgh, PA, USA}
\acmBooktitle{The 37th Annual ACM Symposium on User Interface Software and Technology (UIST '24), October 13--16, 2024, Pittsburgh, PA, USA}
\acmDOI{10.1145/3654777.3676345}
\acmISBN{979-8-4007-0628-8/24/10}

\begin{document}
\title{Improving Steering and Verification in AI-Assisted Data Analysis with Interactive Task Decomposition}

\author{Majeed Kazemitabaar}
\orcid{0000-0001-6118-7938}
\affiliation{%
  \institution{University of Toronto}
  \city{Toronto}
  \state{Ontario}
  \country{Canada}
}
\email{majeed@dgp.toronto.edu}

\author{Jack Williams}
\orcid{0000-0003-1925-7191}
\affiliation{%
  \institution{Microsoft Research}
  \city{Cambridge}
  \country{UK}
}
\email{jack.williams@microsoft.com}

\author{Ian Drosos}
\orcid{0000-0003-3475-2609}
\affiliation{%
  \institution{Microsoft Research}
  \city{Cambridge}
  \country{UK}
}
\email{t-iandrosos@microsoft.com}

\author{Tovi Grossman}
\orcid{0000-0002-0494-5373}
\affiliation{%
  \institution{University of Toronto}
  \city{Toronto}
  \state{Ontario}
  \country{Canada}
}
\email{tovi@dgp.toronto.edu}

\author{Austin Z. Henley}
\orcid{0000-0003-1069-2795}
\affiliation{%
  \institution{Microsoft Research}
  \city{Redmond}
  \state{Washington}
  \country{USA}
}
\email{austinhenley@microsoft.com}

\author{Carina Negreanu}
\orcid{0000-0003-2130-7223}
\affiliation{%
  \institution{Microsoft Research}
  \city{Cambridge}
  \country{UK}
}
\email{cnegreanu@microsoft.com}

\author{Advait Sarkar}
\orcid{0000-0002-5401-3478}
\affiliation{%
  \institution{Microsoft Research}
  \city{Cambridge}
  \country{UK}
}
\email{advait@microsoft.com}

\renewcommand{\shortauthors}{Kazemitabaar, et al.}

\begin{abstract}
LLM-powered tools like ChatGPT Data Analysis, have the potential to help users tackle the challenging task of data analysis programming, which requires expertise in data processing, programming, and statistics. However, our formative study (n=15) uncovered serious challenges in verifying AI-generated results and steering the AI (i.e., guiding the AI system to produce the desired output). We developed two contrasting approaches to address these challenges. The first (\textbf{\incremental}) decomposes the problem into step-by-step subgoals with pairs of editable assumptions and code until task completion, while the second (\textbf{\structural}) decomposes the entire problem into three editable, logical phases: structured input/output assumptions, execution plan, and code. A controlled, within-subjects experiment (n=18) compared these systems against a conversational baseline. Users reported significantly greater control with the \incremental and \structural systems, and found intervention, correction, and verification easier, compared to the baseline. The results suggest design guidelines and trade-offs for AI-assisted data analysis tools.

\end{abstract}

\ccsdesc[500]{Human-centered computing~Natural language interfaces}
\ccsdesc[500]{Human-centered computing~Interactive systems and tools}
\ccsdesc[500]{Human-centered computing~Empirical studies in HCI}

\keywords{Data Analysis, Data Science Assistant, Human-AI Interaction, AI Agents, Generative AI, Large Language Models, Copilot}

\begin{teaserfigure}
  \includegraphics[width=\textwidth]{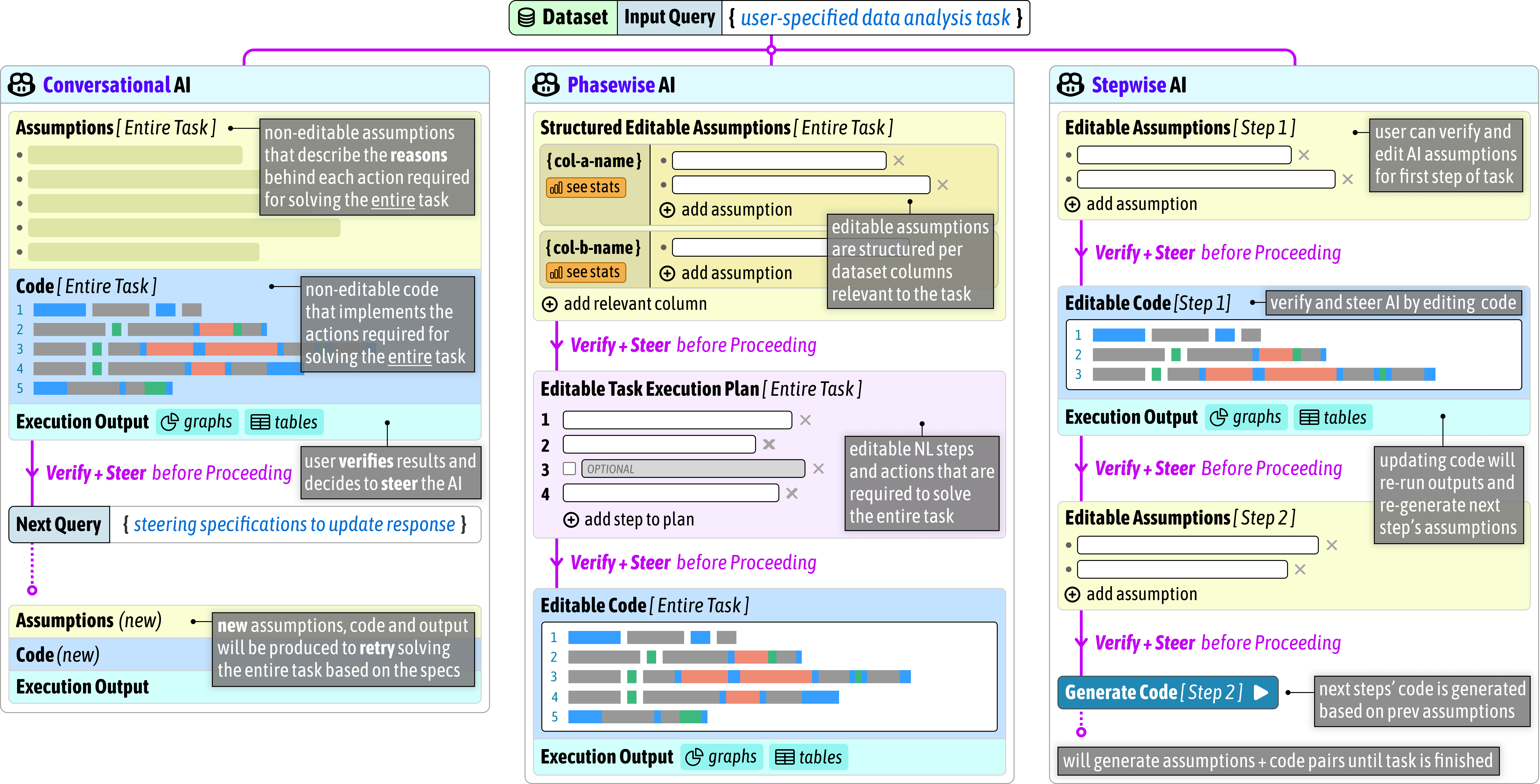}
  \caption{An illustration of the three decomposition approaches that we developed for solving data analysis tasks using AI: (A) \conversational approach solves the entire task without any user intervention but allows submitting follow-up prompts for further steering. (B) \incremental approach provides intervention points at each \textit{step} of solving the task by presenting pairs of editable assumptions followed by corresponding code at each step. (C) \structural approach provides intervention points at each \textit{phase} of solving the entire task with three editable components: editable assumptions of the entire task structured around relevant columns of the dataset, editable task execution plan, and corresponding code.}
    \Description{This illustration is a visual comparison of the three different approaches to steering and verification in AI-assisted data analysis. On the left, the "Conversational AI's Response" section displays a process where the AI generates assumptions and actions in natural language that are not editable, followed by the corresponding code, and finally the execution output, which can be a dataframe, plot, or table. There is only one point indicated for verification and steering the conversation, in the form of asking a follow-up question.
    In the center, the "Stepwise AI's Response" section presents a method where the user can edit assumptions and actions directly. The code is also editable and is followed by an execution output with similar options as the first approach. There are multiple verification and steering points after each step, as well as the ability to generate the next step until the task is completed.
    On the right, the "Phasewise AI's Response" section offers a more structured interaction with editable assumptions and actions linked to specific columns of data, which can be inspected or added to directly. It also features an editable execution plan with numbered steps, where certain steps can be marked as optional. Similar to the other approaches, the code is editable, and the execution output options are the same, with verification and steering points present. Unique to the Phasewise approach is the structured editable input and output for assumptions, which allows users to add assumptions directly to a data column. This system emphasizes a phase-by-phase development where all assumptions, code, and outputs are planned and then executed to solve the full query. Each of these systems has different modalities, with some components being editable and others not, and the number of verification and steering points varies, highlighting the different levels of user engagement and control over the data analysis process.}
  \label{fig:teaser}
\end{teaserfigure}

\maketitle

\section{Introduction}

Data science often involves large datasets, source code, domain expertise, and unwritten assumptions~\cite{muller2019chi}. The process of extracting insights from data~\cite{wang2021autods} for decision making and knowledge discovery~\cite{donoho201750} has several documented challenges~\cite{chattopadhyay2020chi, muller2019chi}. Data scientists spend considerable time inspecting data, writing single-use scripts, ``gluing together'' data sources, cleaning messy data, and documenting their efforts~\cite{chattopadhyay2020chi, kery2017exploring, kery2018story, muller2019chi}. In fact, data scientists describe the need to ``have a conversation'' with their data to understand it~\cite{muller2019chi}.

Recent advancements in AI and particularly the natural language processing and code generation capabilities of Large Language Models (LLMs) have shown promise to facilitate data science tasks. Specifically, chain-of-thought prompting \cite{wei2022chain} and ReAct prompting \cite{yao2022react} have emerged as implementation techniques for \textit{task decomposition}, generating \textit{``reasoning''} traces, followed by \textit{``acting''} traces where the LLM can invoke external agents. In data science, such agents might read from datasets and execute code. For example, ChatGPT Data Analysis supports uploading CSV files, after which its language model and code execution agents can be used to clean data, display visualizations, and answer questions about data through its iterative chat interface~\cite{chatgpt_code_interpreter}. 

However, a study of data scientists using ChatGPT \textit{without} code execution functionality found that participants were unaware of the AI's assumptions when solving the task, found verifying the correctness of results tedious, were overwhelmed by long responses, and could not effectively steer the AI when it made mistakes~\cite{chopra2023conversational}. 

To better understand user behavior, needs, and challenges when performing exploratory data analysis with a conversational AI tool, we conducted a formative study involving 15 participants (Section~\ref{sec:formative_study}). The study identified \emph{steering} and \emph{verification} as the primary limitations of conversational AI tools. \emph{Steering} refers to the user's interaction with the AI to guide its output from an initial state to a desired outcome. \emph{Verification} refers to the user's interaction to understand the AI's output, check its correctness, ensure no incorrect assumptions were used, and decide on further refinement. The study also highlighted the need for new affordances that decompose and display the AI's chain-of-thought reasoning as structured and interactive assumptions, enabling users to modify them at any time, even retroactively. 

Based on this core requirement we developed two systems that make ``decomposition'' a focal point in the interface, not just an implementation detail (Section~\ref{sec:system_design}). First, the \structural system decomposes the problem into three editable phases (assumptions, planning, and code) with increasing levels of specificity. Second, the \incremental system decomposes the task into separate steps (visually similar to a computational notebook), displaying editable assumptions and their corresponding code one step at a time until the task is complete.

Both approaches use the LLM to decompose the task into parts, and help the user focus on one part at a time (as a metacognitive aid~\cite{tankelevitch2023metacognitive}). This enables finer-grained steering than standard conversational prompting~\cite{zamfirescu2023johnny}, and progressive disclosure to reduce information overload~\cite{nielsen2006progressive}. We introduce the idea of using the LLM to generate \emph{editable assumptions} about the input and desired output, based on the task query and data. This also provides a structure for verifying that the AI correctly interpreted the user's intent and translated it into a valid plan. However, when decomposing a task, there is an important space of trade-offs: how much information to display, when to display it, and how many intervention points to provide with how much control. Our two systems occupy different points in this space, and through our user study, we evaluated the trade-offs and identified different situations in which each approach might be beneficial (Sections~\ref{sec:evaluation} and~\ref{sec:results}).

We conducted a controlled, within-subjects experiment (n=18) comparing the \incremental and \structural tools on task decomposition and prompting strategies that support steering, verification, and the user's perceived utility of the tools. We also developed a baseline tool, called \conversational, similar to ChatGPT with code execution. Users reported significantly greater control with the two new systems, and found intervention, correction, and verification easier, compared to the baseline.

This paper makes the following contributions:
\begin{itemize}
    \item A formative study that identifies the limitations of \textit{``conversational''} AI tools in terms of steering them and verifying their output.
    \item A novel approach to improve steering and verification using editable AI assumptions, progressive disclosure, and non-linear conversations to promote data exploration and verification. We describe two implementations of this approach, each balancing information overload and the degree of user control differently (Section~\ref{sec:system_design}).
    \item A within-subjects experiment in which we compared the two systems with a \conversational baseline system (Section~\ref{sec:evaluation}), finding that while there was no difference in task success or completion time, participants felt significantly more in control with the \structural and \incremental systems compared to the baseline (Section~\ref{sec:results}). 
\end{itemize}

\section{Related Work}
\label{sec:related_work}

\subsection{AI-assisted Data Analysis}

Previous work has considered how data transformation scripts can be synthesized from demonstrations (e.g., Wrangler \cite{guo2011proactive, kandel2011wrangler}). This follows an influential line of research that synthesizes programs from examples in data wrangling contexts (e.g., \cite{gulwani2011automating}), which may include natural language \cite{gulwani2014nlyze}. These can be constrained to use specific APIs such as \texttt{pandas}, using generator-based synthesis (e.g., AutoPandas \cite{bavishi2019autopandas}). Scripts can also be synthesized based on heuristics of data quality improvement (e.g, CoWrangler \cite{chopra2023cowrangler}), and data preparation heuristics can also be learned from corpora (e.g., Auto-Suggest \cite{yan2020auto}).

More recently, a number of commercialized LLM supported data analysis tools have become available. These enable data scientists to access AI-powered chat assistants within their notebook (such as Anaconda \cite{anaconda_assistant}, Databricks \cite{databricks_assistant}, and Jupyter AI \cite{jupyter_ai}), and other alternate data-science environments (e.g., DataChat AI \cite{datachat_ai}, SQL and file editors for Databricks, etc.). The semantic abilities of LLMs, coupled with a chat interface, allows conversational interaction with data, follow-up questions, and highly contextualized responses. Consequently, research has investigated the chatbot paradigm for AI assistance in data analysis and visualization in detail \cite{hoon2020interfacing, kassel2018valletto, zhi2020gamebot, setlur2022you, gu2023data, dibia2023lida}.

Thus, early work on data wrangling script synthesis can be contrasted with current LLM-powered data analysis tools both in terms of the complexity of tasks being tackled, and the interaction modality (i.e., from demonstration, examples, and direct manipulation, to naturalistic language prompts). In turn, this also means that generation mistakes become more common, due to underspecification of natural language, assumptions that the AI is making but the user is not aware of, etc. This creates new metacognitive demands for the user to verify the AI's responses and then steer the AI if incorrect. In our work, we try to provide new interaction modalities with LLMs for data analysis tasks to increase the transparency of the AI and the assumptions that it is making.

McNutt et al. \cite{mcnutt2023design} present a design space for AI code assistance in computational notebooks, which are commonly used for data analysis. They find that AI assistants can vary in the gestures they provide for the user to initiate a model response, and options that the user has to verify and refine the output. They also consider the relationship between the assistant interface and other interface components, such as code context, specialization, provenance, and customization.

Though not specifically tackling data analysis, Sarkar et al. \cite{sarkar2022like} studied the experiences of programmers using LLM assistance for writing code. They found that LLM assistance was most useful in rewriting boilerplate code and in API discovery, but also brought new challenges for debugging and code inspection. Sarkar et al. also identify prompt formulation as a major challenge. Fiannaca et al. \cite{fiannaca2023programming} describe methods for how this can be improved by leveraging semantically meaningful structure within prompts to assist programmers.

Similarly, Vaithilingam et al. \cite{vaithilingam2022expectation} found that while programmers preferred LLM-assisted programming to unassisted programming, there were no consistent improvements in task time or success rate, due to productivity benefits being opposed by new challenges in debugging and comprehension of AI-generated code. A detailed telemetric study of GitHub Copilot usage by Mozannar et al. \cite{mozannar2024reading} similarly found that the ``verifying suggestion'' state is the most time consuming. 

There have been other studies of AI-assisted programming \cite{dibia2022aligning, kalliamvakou2022research, liang2023understanding, barke2023grounded}. Many of these point to steering and verification as general challenges with all LLM-assisted programming, which also apply in the specific case of programming data analysis scripts with LLM assistance. 

However, it is worth noting that data analysis does have particularities in comparison to ``general'' programming tasks, e.g., data analysis programming tends to be more exploratory and open-ended, and the activity of analyst sensemaking and insight generation is more important than providing code as a finished product \cite{kery2017exploring, liu2020paths, pirolli2005sensemaking, rule2018exploration, epperson2022strategies, kandel2012enterprise, grolemund2014cognitive, koesten2021talking, russell1993cost}. The implication of this is that the need for rapid steering and verification is more acute in data analysis programming, since the effectiveness of the process depends on rapid exploration of the program space.

\subsection{Verifying LLM Outputs and their Reliability}

Data science is a challenging yet important function within software teams. Previous research has focused on how data scientists engage in collaborative sensemaking, and make choices about how to communicate and report results \cite{kim2016emerging, kim2017data, wang2019data, zhang2020data, crisan2020passing, pang2022data}. They have found that data scientists need support in managing these complex collaborative workflows \cite{kery2018story, kery2019towards, wang2020assessing}. Consequently, research has explored how data scientists can manage, visualize, and trace the evolution of their analysis process \cite{kery2017variolite, head2019managing, wu2020b2, pu2021datamations, xiong2022visualizing}.

Working with an AI assistant may have important differences from a human team. Trust in AI systems is developed differently from trust in human collaborators and is mediated by the conceptual metaphors used to convey them \cite{jung2022great, khadpe2020conceptual}. Trust, communication, and perception management in human collaborations may result in a lack of code verification behaviours, or selective sharing \cite{devito2018people, wang2019data, liu2020paths, pang2022data, mcnutt2023design}. This raises the importance of additional tools for verifying AI generated output, for instance ``co-audit'' tools \cite{gordon2023co}.

Previous research has noted the importance of explainability in AI to support user tasks and decision-making \cite{kim2023help, liao2020questioning, liao2021human}. There are difficulties in applying traditional explainability techniques to large language models due to their large number of parameters, training sets, and complex and open-ended space of inputs and outputs \cite{dibia2022aligning, sun2022investigating, vasconcelos2023generation, sarkar2022explainable}. Research has also explored the challenges non-experts face in prompting and conversational strategies for explainability \cite{ashktorab2019resilient, lakkaraju2022rethinking, zamfirescu2023johnny}.

\highlight{Furthermore, previous research on AI coding assistants emphasizes that engaging users in verifying the \textit{process} of how the AI generates code can maximize user experience, efficiency, and the predictability of obtaining a helpful response \cite{kazemitabaar2024codeAid}. For instance, CodeHelp \cite{liffiton2023codehelp} incorporates a \textit{Sufficiency Check} step that engages users in refining the AI's understanding of the task by prompting them to clarify uncertainties or provide missing context.}

Gu et al. \cite{gu2024analysts} created a design probe similar to ChatGPT Data Analysis \cite{chatgpt_code_interpreter} with an added sidebar for inspecting intermediary variables. They conducted a study with 22 participants using the design probe to understand common behaviors for verifying the AI's response to a natural language query and dataset. They found two main behaviors within the verification workflow: \textit{procedural-oriented} and \textit{data-oriented} which in many cases were tightly coupled and participants frequently switched between an intermediary variable and the code that outputted it. In contrast with how data analysts verify their work in any tool-assisted (non-AI) data analysis workflows, there is now a much bigger demand for verification when users ``offload'' an entire data analysis task to an LLM.

In our work, we explicitly ask the AI to show its assumptions and reasoning in a structured (and editable) way, paired with their corresponding actions, so that users could focus on them and make decisions. We also include features such as ``side queries'' that allow users to pose exploratory questions, build up assumptions, and then add those assumptions to the AI generation workflow.

\subsection{Steering LLMs}
Currently, most commercial LLM tools (e.g., ChatGPT Data Analysis \cite{chatgpt_code_interpreter} or ChatGPT with Noteable \cite{noteable_plugin}) use a turn-based conversational method, \highlight{where the AI attempts to solve problems with minimal intervention points. Typically, users can only steer the AI after it has generated an entire solution, using follow-up prompts, which limits steering control. To address such limitations,} Masson et al. \cite{masson2024directgpt} propose principles of direct manipulation \cite{hutchins1985direct} for steering LLMs in other contexts: continuous representation of objects of interest, physical actions to localize prompt effects, and reusable prompts. Furthermore, research on the metacognitive demands of generative AI identifies decomposition and structured generation as potential aids \cite{tankelevitch2023metacognitive}. Suh et al. \cite{suh2023sensecape} explore hierarchical text generation at different abstraction levels to assist with sensemaking and managing information overload from large text quantities. They also introduce \emph{structured generation} \cite{suh2024structured}, where user's prompt is first used to generate dimensions that make the model's responses vary, and then responses are generated according to those dimensions. 

Specifically in the context of AI-assisted programming, Liu and Sarkar et al. \cite{liu2023wants} introduce ``grounded abstraction matching,'' allowing users to steer LLMs by editing natural language utterances grounded in each step of AI-generated code for data analysis in spreadsheets. \highlight{Similarly, Tian et al. developed \textsc{Steps}, which lets users edit step-by-step explanations of AI-generated SQL code from natural language queries \cite{tian2023interactive}. CoLadder \cite{yen2023coladder} aids experienced programmers in externalizing their problem-solving intentions flexibly, enhancing their ability to evaluate and modify code across various abstraction levels, from goal to final code implementation. These methods enable users to edit natural language prompts grounded in each step of AI-generated code, providing an accessible abstraction level for reading, verifying, and editing.}

\highlight{However, these approaches hide the AI's reasoning and decomposition process, leaving users without insight into the ``how'' and ``why'' behind the generated code. Users are left to manually infer the AI's reasoning from the output and determine explicit actions to edit and refine the grounded utterances. While this might be an acceptable trade-off in systems that generate short programs (e.g., typical spreadsheet formulas or SQL queries), it is unclear how this approach would extend to longer and more complex data analysis scripts. Our work expands this design space by not only displaying the AI's assumptions in a structured way but also enable users to directly edit these assumptions as a novel method of steering the AI to control its output.}

\section{Formative Study}
\label{sec:formative_study}
\highlight{To explore the challenges of data analysis with conversational AI assistants, we conducted a formative study with 15 participants (12 male, 3 female, 0 non-binary) using the Noteable plugin for ChatGPT \cite{noteable_plugin}. At the time of the study, Noteable was the only publicly available tool offering features similar to ChatGPT Data Analysis (formerly Code Interpreter). With Noteable, participants could upload datasets to a Noteable project and enter a natural language (NL) descriptions of their data analysis task in ChatGPT. In response, ChatGPT would generate code cells in the Noteable project, which Noteable would execute. ChatGPT then displayed Noteable's output including any tables or visualizations, and generated an interpretation of the results. ChatGPT would then continue generating code if the task was incomplete, or asked users for additional information if required.}

Participants (F1-F15), who were recruited from our research institute, regularly performed data analysis tasks using computational notebooks and Python data science libraries. Each participant was assigned to one of four tasks commonly performed by data scientists: data cleaning, merging and plotting, extracting insights, or training an ML model (see Table \ref{tab:formative_study_tasks}). 

Study sessions lasted approximately 60 minutes and were conducted in-person. Screen activity was recorded. Participants were asked to think aloud \cite{fonteyn1993description}, and audio data was recorded and transcribed. Participant consent was obtained prior to the study and participants were each compensated with a GBP \textsterling25 Amazon gift card. The study protocol was approved by our institution's ethics and compliance review board.

\begin{table*}[ht]
    \centering
\caption{Tasks used in the formative study in which users used ChatGPT with the Noteable plugin.}
\label{tab:formative_study_tasks}
    \begin{tabular}{ll>{\raggedright\arraybackslash}p{0.4\linewidth}l}
        \toprule
         Task&  Dataset&  Task Description& Assigned Participants\\
         \hline
         Data Cleaning&  \texttt{food-choices.csv}&  Fix columns with inconsistent formatting and prepare dataset for analysis.& F5, F8, F9 \\\\
         Merging and Plotting&  \texttt{country-happiness.csv}&  Merge the five happiness datasets on the country column and visualize the top countries with the moist changes in happiness score from 2015 to 2019.& F10, F11, F12, F13 \\\\
         
 Extracting Insights& \texttt{airbnb-nyc.csv}& Extract high-level spatial and temporal insights about price, availability, and distribution.&F2, F3, F7, F15 \\\\
 Training Model& \texttt{heart-disease.csv}& Train an explainable ML model to predict presence of heart disease and report the key factors contributing to the presence or absence of heart disease in patients.&F1, F4, F6, F14\\
    \bottomrule
    \end{tabular}
\end{table*}

\subsection{Results}
We analyzed the interactions participants had with ChatGPT and the Noteable plugin from 301 total prompts. Participants used a mix of different actions which included (1) directing the AI to perform a data analysis task, (2) exploring the dataset, (3) requesting suggested methods or approaches to accomplish the task, (4) steering and repairing the AI process in how it should accomplish the task, and (5) performing verification on the results of the task (either with or without the AI).

\emph{Steering:} Participants steered the AI's actions and methods using their NL prompts (106 prompts, 35\%). Many of the steering prompts (n=34) were for performing data wrangling (cleaning and manipulation) tasks on specific columns of the dataset. Similarly, some prompts (n=20) were used to explicitly add, remove, or change code produced in previous steps (e.g., \texttt{``exclude the ones that are purely categorical''} (F5)). For repairing mistakes the AI made or any miscommunications between human and AI, participants frequently corrected an assumption the AI had made (36 prompts). For example, after ChatGPT generated data analysis code, F8 prompted ChatGPT that they wanted code that could \texttt{``map each row to multiple classes and not just one closest class''} instead. 

\emph{Data exploration:} We identified 76 instances (25\%) in which participants wanted to inspect the data frames loaded into the notebook using natural language filters such as displaying \texttt{``which country names are inconsistent''} (F12), and \texttt{``unique values in GPA column''} (F8). Sometimes these explorations were in the form of visualizations, (e.g., requesting a \texttt{``histogram of cholesterol levels''} (F6)).

\emph{Verification:} Although the most common behavior for validating the AI's process was reading the AI-generated code and inspecting the output, we also categorized 57 prompts (19\%) as assisting with verification, such as: \texttt{``The USA is missing from all these heat maps, is it also missing from the CSV files or not?''} (F11).

\emph{Code or Logic Explanation:} In 21 prompts (7\%), participants used the main thread of the conversation to ask the AI for explanations about code they did not understand, an algorithm that was used, how something was computed, or help interpreting the results.

\highlight{Furthermore, our results indicate that each participant engaged in linear conversations consisting, on average, 20 AI-generated messages (SD=5). They experienced lengthy responses upon each interaction point, averaging 24 lines of AI-generated code (SD=21, Max=152) and 134 words of the AI's interpretation of the output (SD=104, Max=717). Participants often lost track of the long conversation history and struggled with finding, verifying, and fixing accumulated assumptions. As a result, they requested an ``undo button'' to fix accumulated assumptions made by the AI. Without this feature, they tried workarounds, asking the AI to \texttt{``undo the last step''} (F9) or to \texttt{``ignore the previous data cleansing steps and do it from scratch''} (F5).}

\subsection{Design Goals and Rationale}
Our formative study highlighted \textit{steering} and \textit{verification} as the most common user interactions. Based on our findings, along with relevant prior work and how such AI tools use chain-of-thought prompting for task decomposition and execution, we established the following design goals to enhance user control over the AI-assisted data analysis process.

\highlight{First, participants struggled to understand the AI's reasoning. They often tried to manually infer the underlying assumptions from its generated code, verify them, and then correct them with follow-up prompts. moThis was evident from the 36 prompts that they used to explicitly fix incorrect assumptions made by the AI. Therefore, \emph{\textbf{DG.1} proposes visually separating each different assumption from its corresponding actions (code) and allowing users to directly edit and update them.}}

\highlight{Second, participants were overwhelmed by long responses and lost track of the long conversation history. Consistent with previous studies \cite{chopra2023conversational}, \emph{\textbf{DG.2} recommends adding intervention points in the AI's responses to help users focus on smaller information chunks. Additionally, steering operations should update only relevant sections at each intervention point, rather than adding new outputs to the main thread.}}

\highlight{Lastly, participants frequently used prompts for data exploration, result verification, and code explanation. While these were useful, they often derailed the main conversation thread from solving the task. To address this, \emph{\textbf{DG.3} suggests enabling side conversations and other methods to assist users in verifying assumptions without cluttering the main thread.}}

\section{System Design}
\label{sec:system_design}
We address design goal \highlight{\textbf{DG.1}} through \textit{interactive task decomposition}. This involves: (1) prompting the LLM to generate its chain-of-thought reasoning as NL assumptions and corresponding actions about the input task and dataset; and (2) rendering the LLM's output as structured, editable UI components, allowing users to refine the AI's proposed plan. We refer to these as \textit{editable assumptions} \highlight{that represent the AI's reasoning based on the task and dataset (e.g., pattern of values in a dataset column.)}

In addressing \highlight{\textbf{DG.2}} to provide proper intervention points, we encounter trade-offs in balancing the number, amount of information presented, and the degree of control provided at each intervention point. This led us to develop two alternative approaches. The \textbf{\structural} system, which gives users greater control over the entire analysis plan from the outset, but with fewer intervention points, and requires the user to understand more information at each step. Conversely, the \textbf{\incremental} allows more focused control by decomposing the task into step-by-step subgoals, increasing intervention opportunities, reducing the information overload per step, but with less structured control over the entire task.

Similarly, to address \highlight{\textbf{DG.3}}, we balanced the amount of information displayed for verification and decision making at each intervention point. The \structural system aids AI-based information retrieval by displaying relevant dataset columns, allowing inspection of column statistics and AI assumptions, as well as distinguishing required and suggested steps in the execution plan. To support user-led exploration in both systems, we allocated a ``sidebar'' on the ride side of the screen where users can: (a) select portions of the AI-generated code and ask questions about them; (b) ask natural language queries for data exploration; and (c) generate code from natural language description for the user to manually incorporate into the AI-generated code. 

In the following, sections we outline the core features shared between the \structural and \incremental systems, and explain how they differ. Lastly, we describe the \conversational system, serving as a baseline similar to ChatGPT's Advanced Data Analysis plugin for our user evaluation.

\subsection{Core System Features}~\label{sec:core_system_features}
\begin{figure}
    \centering
    \includegraphics[width=1\linewidth]{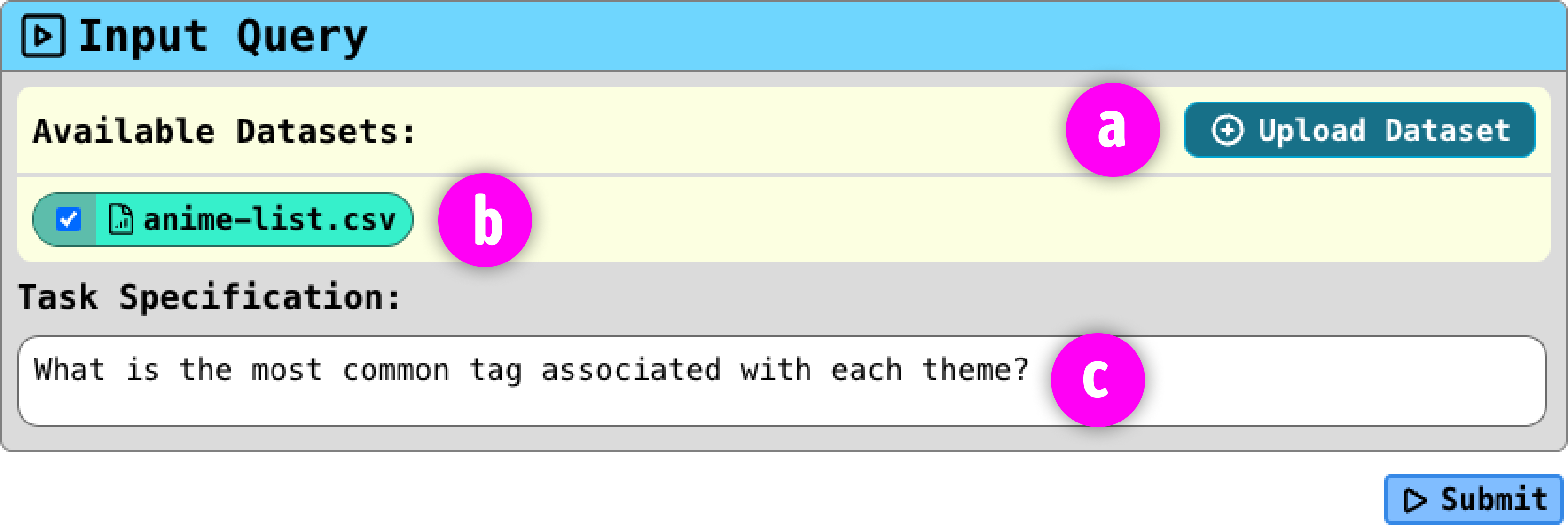}
    \caption{The starting input for all systems, \textmd{which includes a button to upload datasets \protect\raisebox{-1.8pt}{\includegraphics[scale=0.17]{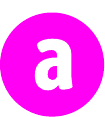}}, selectable datasets to be included in the analysis \protect\raisebox{-1.8pt}{\includegraphics[scale=0.17]{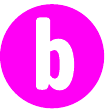}}, and a text input for entering the natural language description that specifies the data analysis task \protect\raisebox{-1.8pt}{\includegraphics[scale=0.17]{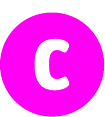}}.}}
    \label{fig:shared_input_query}
\end{figure}

\begin{figure}[!htb]
    \centering
    \includegraphics[width=1\linewidth]{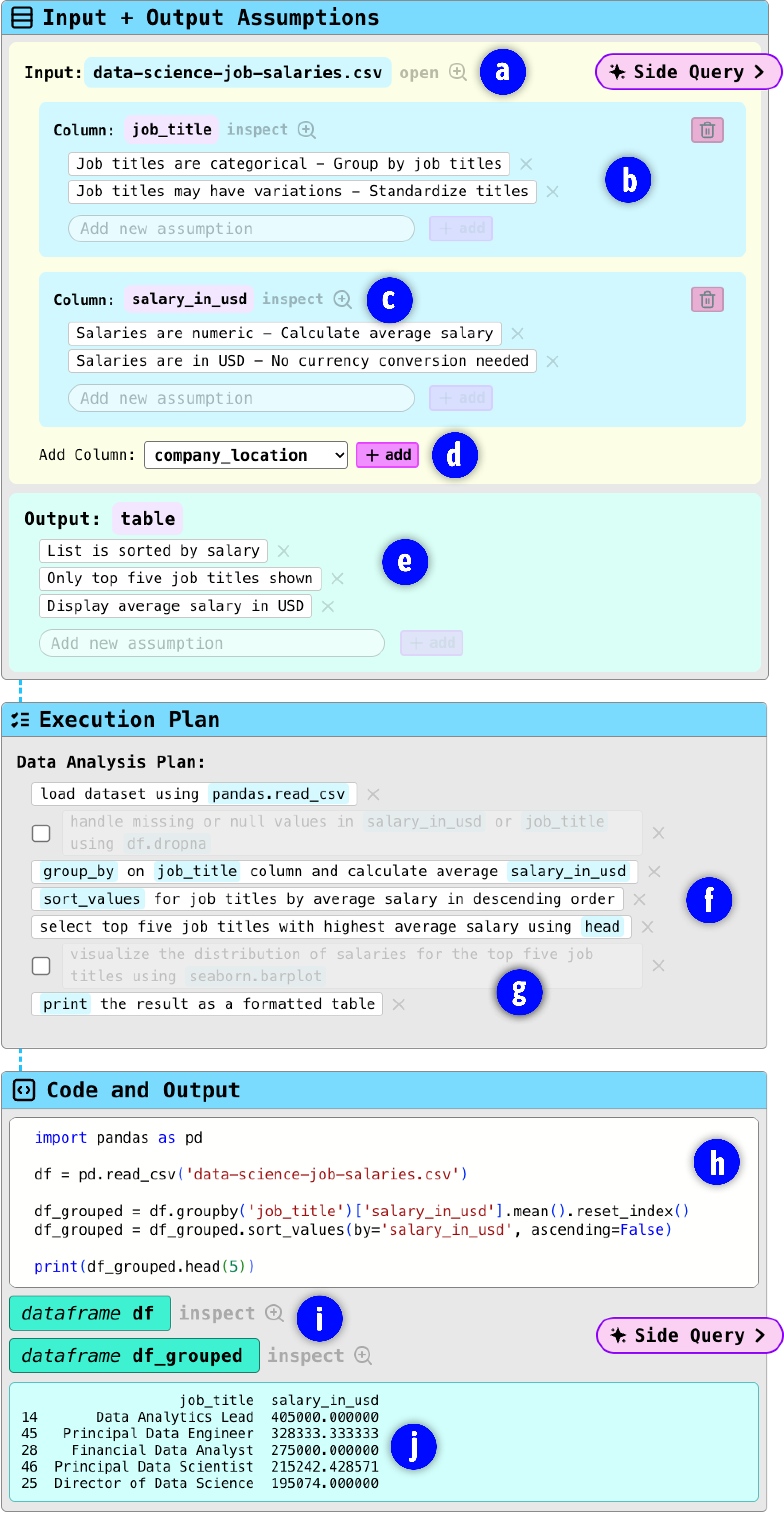}
    \caption{Overview of the \structural system's task flow, \textmd{which decomposes tasks into three stages. \textit{Input + Output Assumptions}, allows users to upload a dataset \protect\raisebox{-1.8pt}{\includegraphics[scale=0.17]{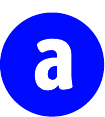}}, manage column-based AI assumptions \protect\raisebox{-1.8pt}{\includegraphics[scale=0.17]{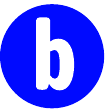}}, inspect column-based descriptive statistics \protect\raisebox{-1.8pt}{\includegraphics[scale=0.17]{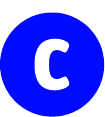}}, add columns missed by the AI \protect\raisebox{-1.8pt}{\includegraphics[scale=0.17]{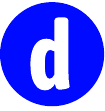}}, and edit assumptions about the task's output \protect\raisebox{-1.8pt}{\includegraphics[scale=0.17]{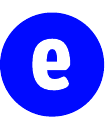}}. \textit{Execution Plan} contains the AI's editable natural language plan for solving the task \protect\raisebox{-1.8pt}{\includegraphics[scale=0.17]{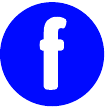}}, which includes user-selectable optional steps \protect\raisebox{-1.8pt}{\includegraphics[scale=0.17]{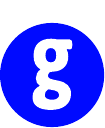}}. \textit{Code and Output} contains AI generated code for solving the task and includes an code editor \protect\raisebox{-1.8pt}{\includegraphics[scale=0.17]{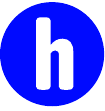}}, intermediary variable inspector \protect\raisebox{-1.8pt}{\includegraphics[scale=0.17]{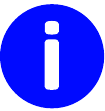}}, and the code output \protect\raisebox{-1.8pt}{\includegraphics[scale=0.17]{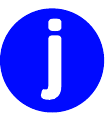}}. Section \ref{sec:taskdecompstruc} details these features.}}
    \label{fig:comprehensive_system_screenshot}
\end{figure}

\subsubsection{Task Input}
Data analysis begins with dataset(s) and a task specification.
\emph{Dataset Input:} Data is loaded using the \textit{Input Query} interface (Figure \ref{fig:shared_input_query}, \protect\raisebox{-1.8pt}{\includegraphics[scale=0.17]{figures/ovals/pink-a.pdf}}). Users can then select one or more datasets relevant to the task (Figure \ref{fig:shared_input_query}, \protect\raisebox{-1.8pt}{\includegraphics[scale=0.17]{figures/ovals/pink-b.pdf}}). The Python server generates a summary of each selected dataset with sample values for all columns. This summary is passed to the LLM to build its initial set of assumptions for the data analysis task.

\emph{Task Specification Input:} After selecting relevant datasets, users specify their data analysis task in the text field and press submit to start the data analysis process (Figure \ref{fig:shared_input_query}, \protect\raisebox{-1.8pt}{\includegraphics[scale=0.17]{figures/ovals/pink-c.pdf}}).

\subsubsection{Task Decomposition: \structural System}
\label{sec:taskdecompstruc}
Using the summary of the dataset and the user-specified task description, the \structural system decomposes the task into three phases: Input and Output Assumptions, Execution Plan, and Code and Output. 

\emph{\textbf{A)} Input and Output Assumptions:} After loading a dataset (Figure \ref{fig:comprehensive_system_screenshot}, \protect\raisebox{-1.8pt}{\includegraphics[scale=0.17]{figures/ovals/blue-a.pdf}}), this component displays all the columns that the AI found to be relevant to the task, and for each column it displays several editable assumptions regarding the task (Figure \ref{fig:comprehensive_system_screenshot}, \protect\raisebox{-1.8pt}{\includegraphics[scale=0.17]{figures/ovals/blue-b.pdf}}).
Assumptions can pertain to data type, uniformity, units, sorting order, etc.
Users can delete columns they find unrelated to the task, or add columns that the AI incorrectly did not select (Figure \ref{fig:comprehensive_system_screenshot}, \protect\raisebox{-1.8pt}{\includegraphics[scale=0.17]{figures/ovals/blue-d.pdf}}). Within each column, users can edit, add, or remove assumptions for that column (Figure \ref{fig:comprehensive_system_screenshot}, \protect\raisebox{-1.8pt}{\includegraphics[scale=0.17]{figures/ovals/blue-b.pdf}}).
For each column, users can ``inspect'' descriptive statistics (Figure \ref{fig:comprehensive_system_screenshot}, \protect\raisebox{-1.8pt}{\includegraphics[scale=0.17]{figures/ovals/blue-c.pdf}}), including a frequency table of sample values for categorical columns. 
Additionally, the entire dataset can be viewed by clicking on the ``open'' button \protect\raisebox{-1.8pt}{\includegraphics[scale=0.17]{figures/ovals/blue-a.pdf}}, with the selected columns highlighted to help the user leverage the columns to build up assumptions.
Finally, the task's output assumptions can be viewed and changed to edit, add, or remove assumptions to steer the final output (Figure \ref{fig:comprehensive_system_screenshot}, \protect\raisebox{-1.8pt}{\includegraphics[scale=0.17]{figures/ovals/blue-e.pdf}}).

\emph{\textbf{B)} Execution Plan:} Using the assumptions, including edits, the system generates a list of natural language steps for solving the task (Figure \ref{fig:comprehensive_system_screenshot}, \protect\raisebox{-1.8pt}{\includegraphics[scale=0.17]{figures/ovals/blue-f.pdf}}). Steps are editable and the user may add or remove steps. The model is also prompted to include optional steps that are rendered as selectable steps with a checkbox (Figure \ref{fig:comprehensive_system_screenshot}, \protect\raisebox{-1.8pt}{\includegraphics[scale=0.17]{figures/ovals/blue-g.pdf}}). After the user is satisfied with the plan, they can proceed to generating and running the code.

\emph{\textbf{C)} Code and Output:} Here the AI generates code to solve the task based on the previous two components. The code is immediately executed and displayed in an editor to allow modification and re-execution (Figure \ref{fig:comprehensive_system_screenshot}, \protect\raisebox{-1.8pt}{\includegraphics[scale=0.17]{figures/ovals/blue-h.pdf}}). Users can inspect the dataframe and variables used in the code execution (Figure \ref{fig:comprehensive_system_screenshot}, \protect\raisebox{-1.8pt}{\includegraphics[scale=0.17]{figures/ovals/blue-i.pdf}}), and see the code output (Figure \ref{fig:comprehensive_system_screenshot}, \protect\raisebox{-1.8pt}{\includegraphics[scale=0.17]{figures/ovals/blue-j.pdf}})

\subsubsection{Task Decomposition: \incremental System}
\label{sec:taskdecompincremental}
\begin{figure}[!ht]
    \centering
    \includegraphics[width=1\linewidth]{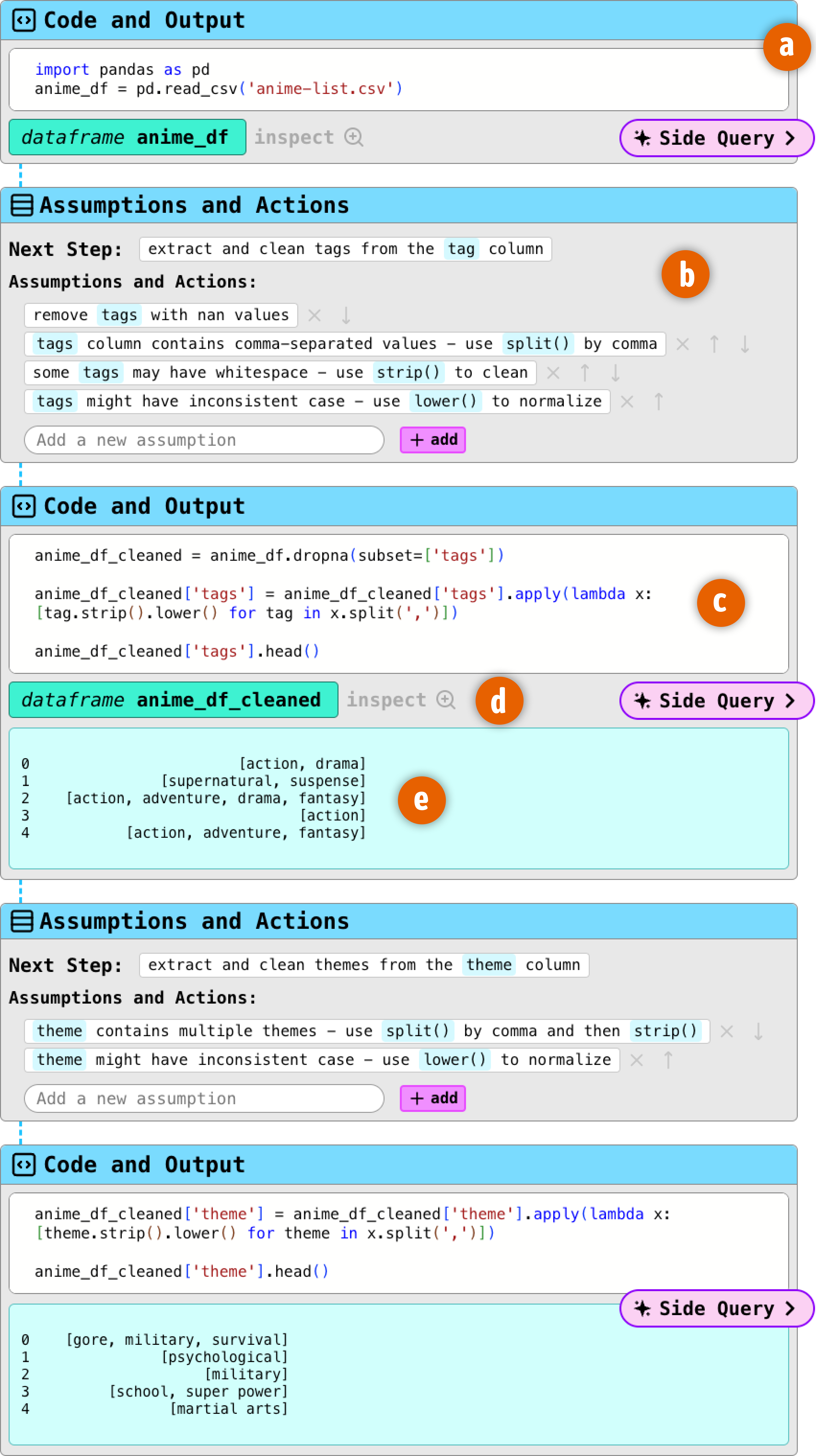}
    \caption{Overview of the \incremental system's task flow, \textmd{which decomposes each task into subgoals containing two components. \textit{Assumptions and Actions} includes the NL subgoal and editable AI assumptions \protect\raisebox{-1.8pt}{\includegraphics[scale=0.17]{figures/ovals/orange-b.pdf}}. \textit{Code and Output} contains a code editor \protect\raisebox{-1.8pt}{\includegraphics[scale=0.17]{figures/ovals/orange-a.pdf}} \protect\raisebox{-1.8pt}{\includegraphics[scale=0.17]{figures/ovals/orange-c.pdf}}, a dataframe and variable inspector \protect\raisebox{-1.8pt}{\includegraphics[scale=0.17]{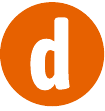}}, and the code output \protect\raisebox{-1.8pt}{\includegraphics[scale=0.17]{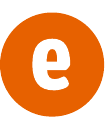}}. Section \ref{sec:taskdecompincremental} details these features.}}
    \label{fig:incremental_system_screenshot}
\end{figure}

Unlike the \structural system where each component reflects the entire task, the \incremental system decomposes the task into subgoals, which have intermediate objectives. Each subgoal (except the first, which loads the dataset (Figure \ref{fig:incremental_system_screenshot}, \protect\raisebox{-1.8pt}{\includegraphics[scale=0.17]{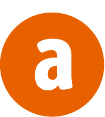}})), is represented as a pair of components: Subgoal Assumptions and Actions, and Subgoal Code and Output.

\emph{\textbf{A)} Subgoal Assumptions and Actions:} Each subgoal starts with a short description of its objective in natural language, followed by several assumptions and actions based on the dataset or previous steps (Figure \ref{fig:incremental_system_screenshot}, \protect\raisebox{-1.8pt}{\includegraphics[scale=0.17]{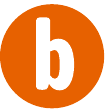}}). 
We designed LLM prompts so that each subgoal would focus on one specific objective such as pre-processing data, filtering columns, performing calculations, and displaying plots. Users may reorder assumptions and actions \highlight{to change their priority}, add or remove assumptions, and edit them directly. Once the user is satisfied with them, they can proceed to generate the subgoal code and output.

\emph{\textbf{B)} Subgoal Code and Output:} Similar to Code and Output in the \structural system, in this component, the system generates code to solve the task based on the previous assumptions and actions (Figure \ref{fig:incremental_system_screenshot}, \protect\raisebox{-1.8pt}{\includegraphics[scale=0.17]{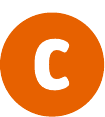}}). The code is immediately executed and can be edited. Once executed, the system generates the next subgoal to allow the user to either reflect on the current subgoal or start working on the next. This process continues until the task is finished and the requirements have been satisfied, in which case the next \textit{Subgoal Assumptions and Actions} will indicate completion. However, if the user still wants to continue, they can add assumptions and actions to continue.

\subsubsection{Editable LLM Assumptions and Actions}

We prompted the LLM to generate each assumption paired with its corresponding action in the format of \inlinecode{<assumption> - <action>}. We also prompted the LLM to enclose column names, variables, and keywords in backticks, which could be rendered into editable components highlighted with a different color. The aim of these interventions was to reduce information overload and improve the efficiency of the editing process.


%

\subsubsection{Code Execution and Intermediary Variables}
The Python server runs the AI-generated code and returns any outputs, including text, visualization plots, or any errors. Any variables and dataframes created during execution are displayed as intermediary variables that the user may inspect.

\emph{Inspect Intermediary Variables:} users can click on each intermediary variable to open a full-screen window for inspecting its values. For dataframes, the interface includes a string matching filter to assist users in finding specific values.

\begin{figure}[!ht]
    \centering
    \includegraphics[width=1\linewidth]{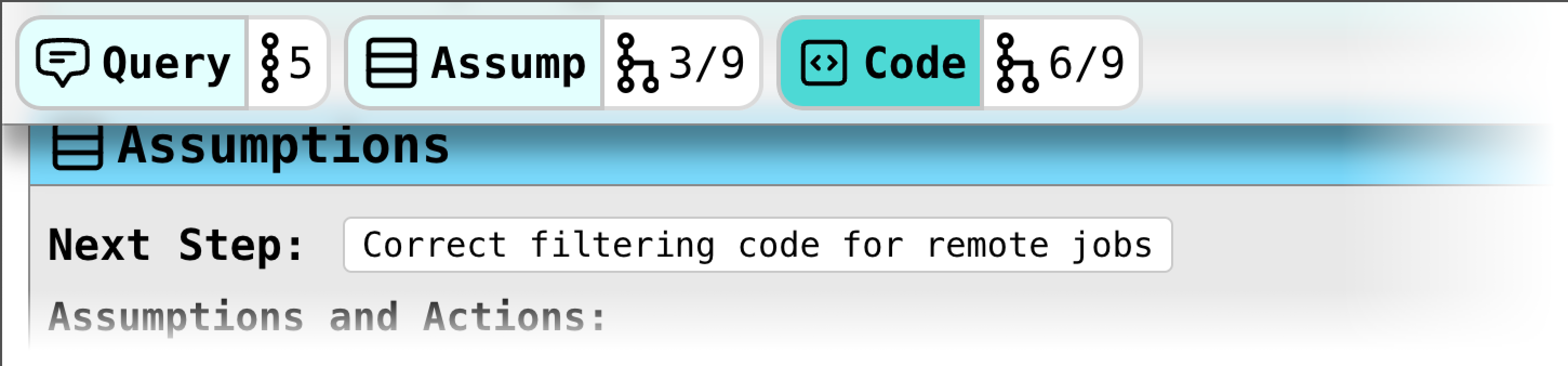}
    \caption{The tabbed ribbon displays all the branches created after editing various nodes. Users can select a different tab to switch to that branch. Each tab indicates where the edit occurred and how much it has progressed (number of total nodes).}
    \label{fig:tabbed_ribbon}
\end{figure}

\subsubsection{Managing Edits}
Within each component, edits can be either pending or submitted. A submitted edit means that the edits have been applied to either generate the next component or regenerate downstream components, whereas a pending edit has not. Pending edits can be reverted using an undo button. However, once an edit is submitted, our system introduces a new branch to preserve the original, unedited version, while incorporating the edited version in the ``main'' branch. New branches are displayed in a tabbed ribbon at the top of the UI \highlight{as shown in Figure \ref{fig:tabbed_ribbon}}. To allow iteration while minimizing proliferation of branches, new branches are not created when the user edits the last generated component in the stream of components. Branching allows users to keep track of previous edits and switch between edits as needed.

\begin{figure}[!ht]
    \centering
    \includegraphics[width=1\linewidth]{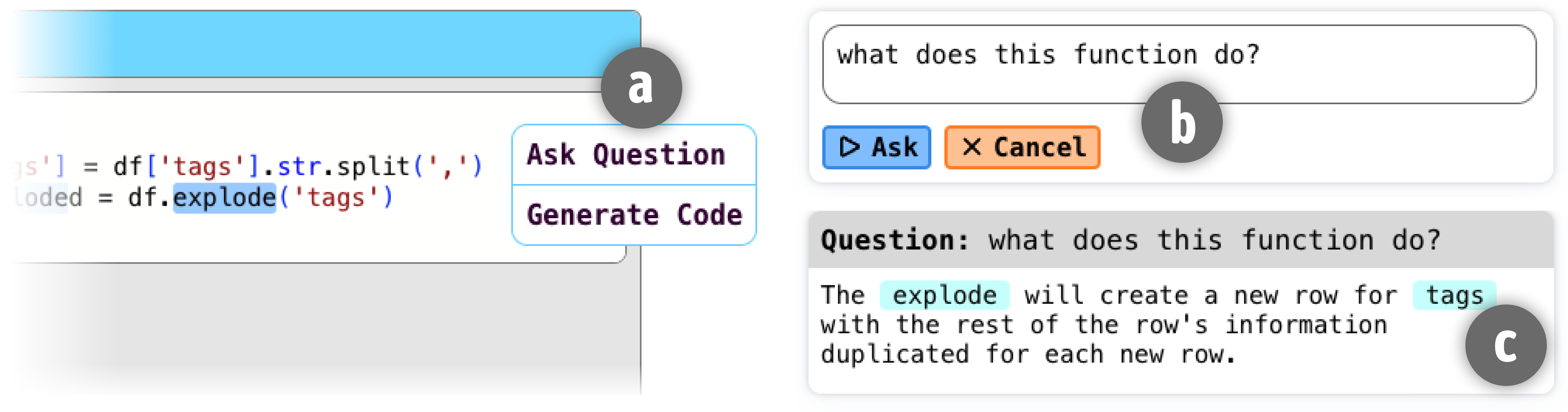}
    \caption{In the \incremental and \structural systems, users can select any code in the editor to ask questions \protect\raisebox{-1.8pt}{\includegraphics[scale=0.17]{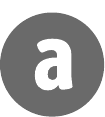}} from the AI. This will create a question box to the right of the main components in which users can ask their clarification question \protect\raisebox{-1.8pt}{\includegraphics[scale=0.17]{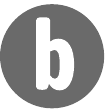}}. The question box will then be replaced with the AI's response \protect\raisebox{-1.8pt}{\includegraphics[scale=0.17]{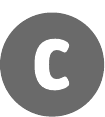}}, based on the query and the selected text.}
    \label{fig:ask_question_screenshot}
\end{figure}

\subsubsection{Side Conversations}
We allocated space to the right of the main components for running side conversations with the system in three formats: \textit{Ask Question}, \textit{Generate Code}, and \textit{Run Side Query}. These features are available in all editable code execution blocks, with the exception of \textit{Run Side Query}, which is also accessible alongside the Input and Output Assumptions in the \structural AI system.

\emph{Ask Question:} This allows users to ask questions about the generated code (See Figure \ref{fig:ask_question_screenshot}). When a code editor is in focus or code is selected, the Ask Question button appears to the right of the editor. The user can provide a natural language query and the system generates a response on the side. The selection allows users to ask targeted questions such as \texttt{``what does this function do?''}

\emph{Generate Code:} This feature generates code based on the selected code segment and the user's query. The user can inspect the generated code and, if it is found useful, insert it into the editor. Similar to the \textit{Ask Question} feature, the selection here enables asking targeted questions, such as updating the code to exhibit a different behavior based on a natural language prompt.

\begin{figure}[!ht]
    \centering
    \includegraphics[width=1\linewidth]{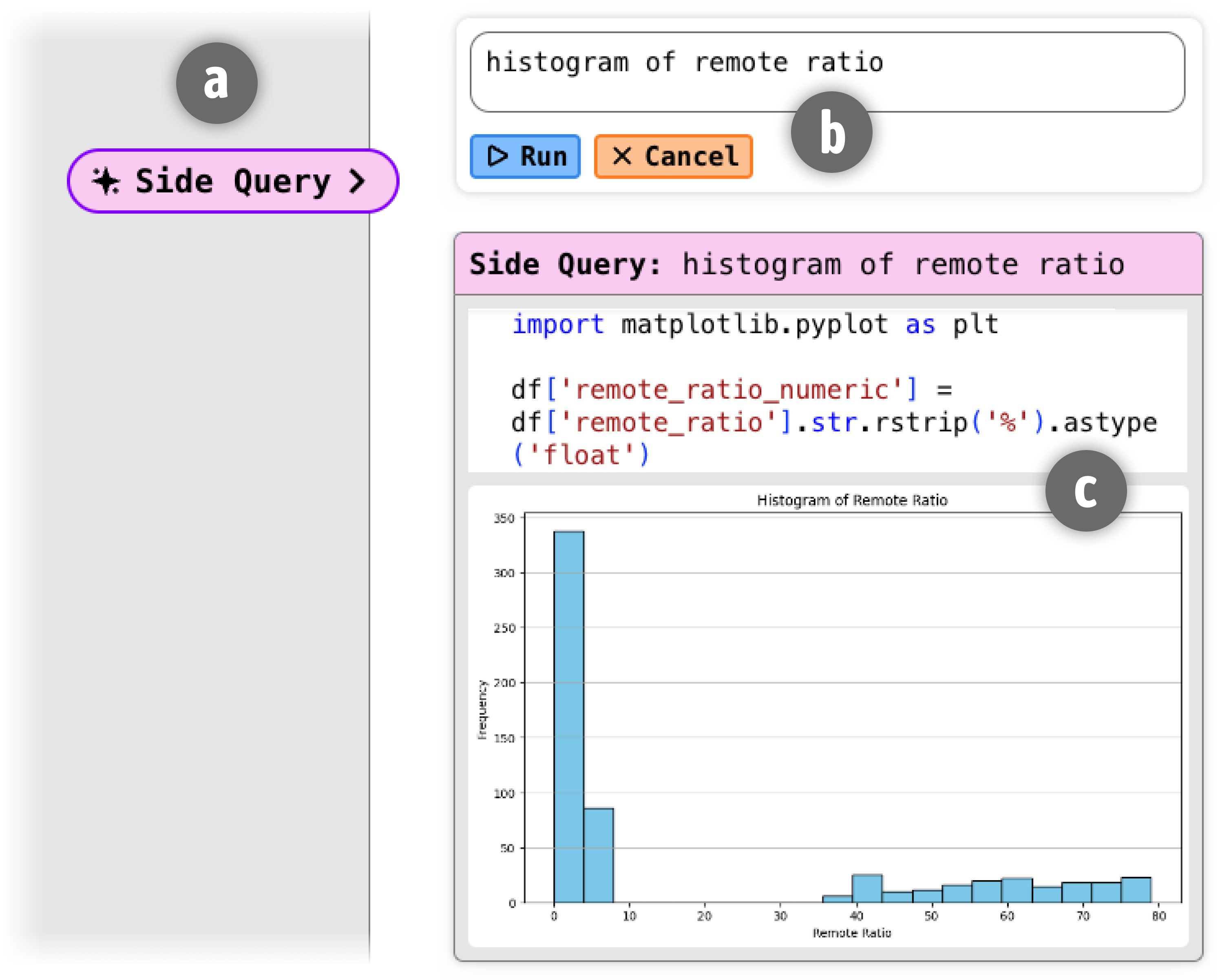}
    \caption{The run Side Query button \protect\raisebox{-1.8pt}{\includegraphics[scale=0.17]{figures/ovals/gray-a.pdf}} is placed to the right of each code execution block in the \incremental and \structural systems, and the input and output assumptions in the \structural system. This will open the question box \protect\raisebox{-1.8pt}{\includegraphics[scale=0.17]{figures/ovals/gray-b.pdf}} in which the user can specify a natural language data exploration query on any of the intermediary variables or the original dataset. The question box will then be replaced with the AI's response which includes the code and any outputs, including visualizations \protect\raisebox{-1.8pt}{\includegraphics[scale=0.17]{figures/ovals/gray-c.pdf}}.}
    \label{fig:side_query_screenshot}
\end{figure}

\emph{Run Side Query:} This feature enables ancillary data analysis tasks using natural language queries. It enables further exploration of the dataset or any intermediary dataframes, and helps users validate and refine assumptions. By clicking on the Side Query button (Figure \ref{fig:side_query_screenshot}), users can ask natural language queries about the dataset or the current state of the code and variables. The system generates and executes code in the side panel, allowing users to view outputs such as visualizations, identifying outliers, and check the data's consistency.

\subsection{\conversational Baseline System}
We developed a \conversational system similar to ChatGPT's Advanced Data Analysis plugin as a baseline to compare with the \structural and \incremental systems. The \conversational system does not include any intervention points or editable assumptions, or any of the side conversation features (e.g. \textit{Ask Question} or \textit{Run Side Query}). It decomposes the task into a bullet point of non-editable, natural language assumptions and actions about the task, and then immediately generates and runs non-editable code that solves the entire task. To interact with this system, as with ChatGPT, the user needs to issue follow-up prompts. In this baseline system only the prompts (and follow-up prompts) are editable. For verification, users could read the code and inspect the intermediate variables, and for steering, they could ask follow-up questions in natural language.

\subsection{System Implementation}
All three variants are built as a web application and Python server stack. The web application is written in TypeScript and the React web framework to include the interface elements described in Section \ref{sec:core_system_features}. The web application interacted with the Python server for uploading datasets, obtaining their descriptive summaries, running code, and retrieving their execution results. It also called the GPT-4 Turbo models from OpenAI. Each component in the \structural or \incremental system is represented as a node in a tree data structure inside the application. This enables tracing the path from each node to the root node to prepare the context prompt for interacting with the LLM and generating the next component. It also provides state management for the edits that create branches and is used to render the tabbed ribbon interface. The Monaco Editor is used as the code editor in each of the code execution blocks and for syntax highlighting the non-editable code pieces. To enable code execution on the Python server, and to retrieve code execution outputs and intermediary variables, we used the IPython kernel. We used the \inlinecode{\%matplotlib inline} command which returned all plots as base64 images that could be included as a response inside the REST APIs.

\begin{figure}[!ht]
    \centering
    \includegraphics[width=1\linewidth]{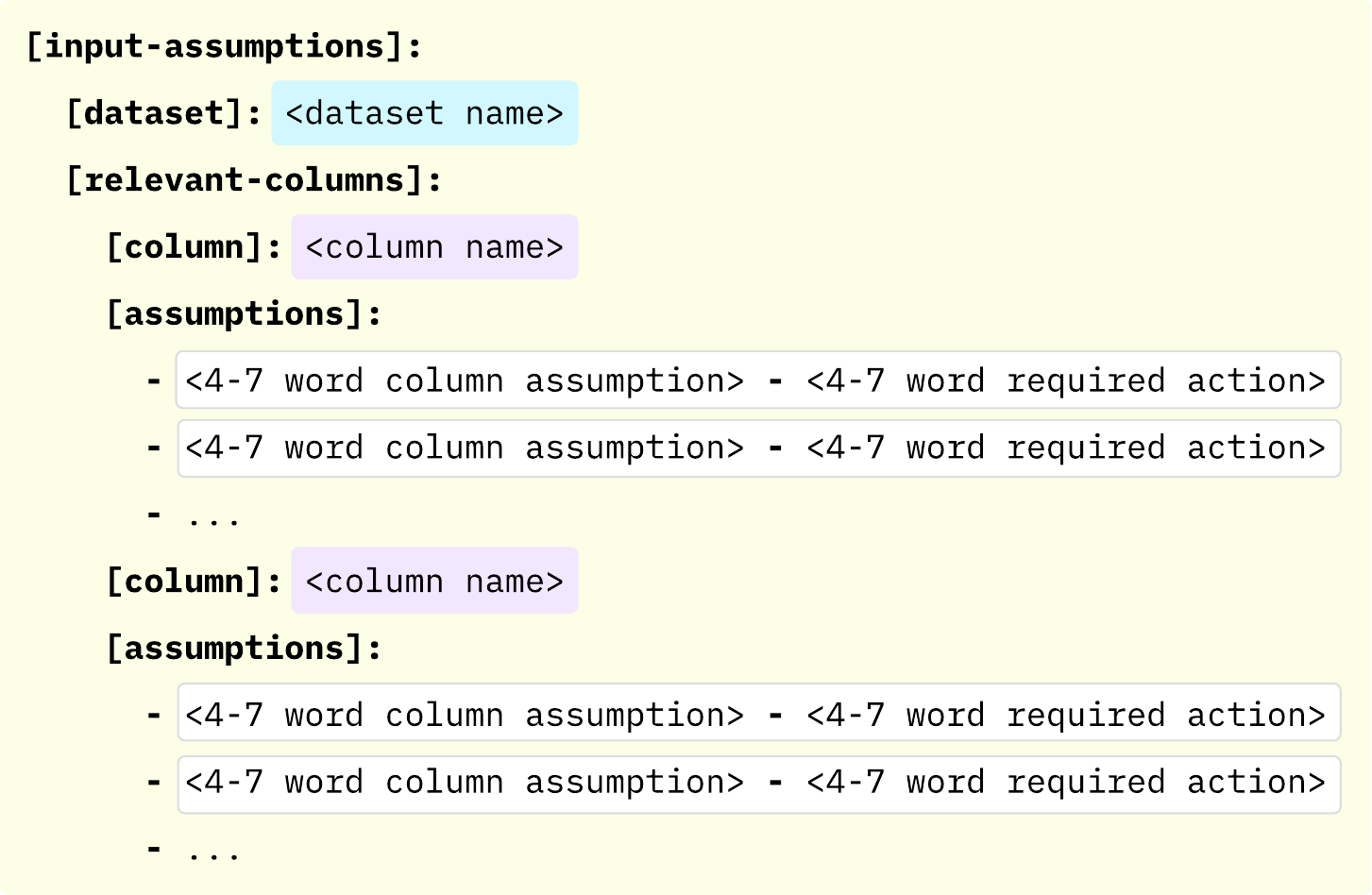}
    \caption{The prompt template that selects dataset columns relevant to the task and generates assumptions and actions for those columns in the \structural AI system.}
    \label{fig:prompt_template}
\end{figure}

\subsubsection{LLM Prompt Structures}
We required complete control over the format of the LLM's output to allow reliable parsing and rendering of structured components. However, few-shot learning (e.g. providing specific input and output examples) would make the LLM overfit to the provided few-shot examples. Through informal experimentation with different prompts and models, we concluded that the GPT-4 and GPT-4 Turbo models are capable of following templates that only specify the format of the output with minimal specification of the content to be generated, with sufficient reliability for a practical evaluation. 
Figure \ref{fig:prompt_template} shows an example of the prompt used to select the columns relevant to the task and generate assumptions and actions about each column. Although the exact format and structure is explicitly provided, the values are not, which enables the system to work generally on a variety of input tasks and datasets.

\section{User Evaluation}
\label{sec:evaluation}
To evaluate and compare the \incremental and \structural systems in enabling users to steer the AI and verify its responses, we conducted a within-subjects study. The study compared these systems with the \conversational baseline and involved 18 participants who used all three systems to complete six data analysis tasks, with two tasks per system. Datasets and tasks were designed to be sufficiently complex that the AI would not automatically produce correct solutions without user involvement. They required participants to carefully verify the AI's process and responses and steer the AI in addressing any issues.

The main focus of our exploratory study is on understanding the unique ways in which each system aids in steering and verification during the AI-assisted data analysis process. We also investigated the perceived utility of other various system components, and explored the usage patterns and user preferences that emerged with each system.

\subsection{Participants}
We recruited 18 participants (10 men, 8 women, 0 non-binary) from a large research university. Participants were pre-screened to ensure they were proficient in writing Python code, familiar with Python data science libraries, and experienced in regularly performing data analysis tasks. In terms of data analysis experience, five participants reported having 1--2 years, seven having 3--5, and six more than five years. The majority (14 participants) used Python daily, while the rest used it at least weekly. All reported familiarity with data science libraries like \texttt{numpy}, \texttt{matplotlib}, with 15 also familiar with \texttt{pandas}. Jupyter Notebooks were used daily by eight participants, weekly by six, monthly by two, and rarely by two. For English proficiency in technical contexts, 16 participants felt very comfortable, while two felt somewhat comfortable. In LLM usage for coding, seven reported using daily, seven weekly, and four using monthly or less.

\subsection{Data Analysis Tasks}
We designed six tasks derived from the ARCADE benchmark \cite{yin2022natural}, which contains a diverse set of tasks from various datasets on Kaggle \cite{kaggle_website}. These tasks included a series of natural language (NL) queries written by professional data scientists with the intention of interacting with an AI assistant. We selected tasks to not require specific domain knowledge, targeting for participants to solve them within a 15-minute time frame. However, we also wanted to make sure that the tasks included additional complexities that would make it difficult for the AI to correctly solve them without proper user verification and intervention. Therefore, we selected and altered tasks and their corresponding datasets with some format inconsistencies and altered distributions. For instance, Task 2 (Table \ref{tab:final_study_tasks}) requires splitting tags and themes with commas before grouping tags by themes. To increase complexity, we modified the tags and themes columns to have only the first theme or tag in capital case, with the rest in lowercase. See Table \ref{tab:final_study_tasks} for details of the six study tasks. We ran each of the six tasks (query + dataset) 10 times on all three systems, ensuring a consistent 100\% failure rate.

\begin{table*}
    \centering
\caption{Final tasks used in the evaluation, including the exact queries for each task, the datasets involved, and issues the AI would encounter without user intervention.}
\label{tab:final_study_tasks}
    \begin{tabular}{>{\raggedright\arraybackslash}p{0.3\linewidth}>{\raggedright\arraybackslash}p{0.25\linewidth}>{\raggedright\arraybackslash}p{0.36\linewidth}}
    \toprule
         Natural Language Query&  Dataset& Issues\\
         \hline
         \textbf{Task 1}: Show me the top five highly rated products by Nivea&  \texttt{big-basket-products.csv}& Not recognizing multiple sub-brands of \inlinecode{"Nivea"} (1) and not cleaning the \inlinecode{Rating} column to accurately extract numeric values (2).\\\\
         \textbf{Task 2}: What is the most common tag associated with each theme?&  \texttt{anime-list.csv}& Not correctly splitting \inlinecode{Themes} by comma (1) and overlooking the inconsistency in casing between \inlinecode{Tags} and \inlinecode{Themes} (2).\\\\
         \textbf{Task 3}: Display the top 20 most popular drama names that have only one unique genre? Popularity is based on drama rating and votes.&  \texttt{korean-drama.csv}& Not filtering \inlinecode{genres} labeled as \inlinecode{"Unknown"} (1) and extreme outliers in \inlinecode{votes} (2)\\\\
         \textbf{Task 4}: What are the top ten positions (based on mean salary) for working remotely in US-based companies?&  \texttt{data-science-job-salaries.csv}& Not cleaning \inlinecode{Country Code} (1) and not identifying remote companies using \inlinecode{Remote Ratio} (2).\\\\
         \textbf{Task 5}: Show the top five movies with the highest percentage return on investment.&  \texttt{bollywood-movies.csv}& Failed to (1) clean \inlinecode{budget} correctly, and (2) calculate missing \inlinecode{Revenue} values based on \inlinecode{India} and \inlinecode{Worldwide}.\\\\
         \textbf{Task 6}: What were the top three lowest scoring matches? Sort in ascending order and show location, local and visitor team names.&  \texttt{euroleague-basketball.csv}& Failed to select related columns for calculating scores (1), and not knowing how to aggregate scores by \inlinecode{Game} and \inlinecode{Round} (2).\\
         \bottomrule
    \end{tabular}
\end{table*}

\subsection{Study Procedure}
The order of the three \structural, \incremental, and \conversational (baseline) systems was \highlight{fully counterbalanced using a Latin square design across the 18 participants and tasks to minimize order effects}, while tasks were fixed from T1 to T6. Participants spent approximately 50 minutes with each system. \highlight{They received a 10-minute tutorial and a 5-minute warm-up task to familiarize themselves with the system. Then proceeded to the main study tasks, where they were given a dataset and a NL query. The lead author who conducted all experiments,} explained the dataset and relevant columns for each task. Participants proceeded to execute the task and were asked to think aloud throughout the study \cite{fonteyn1993description}.

Participants were made aware that identifying and correcting mistakes made by the AI was their responsibility. They were instructed to notify the experimenter once they believed they have achieved a correct result using the AI tool. \highlight{The experimenter would then verify their solution against expected outcomes and provide up to two hints if necessary. These hints addressed AI mistakes correlating with each of the two issues for each task, as listed in Table \ref{tab:final_study_tasks}, ensuring consistency across participants.} Completion criteria for each task required resolving both issues listed in the table. 

Following the completion of each task, participants were asked about their choice of method for steering the AI (e.g., editing the execution plan versus directly editing the code) and their verification processes. After completing two tasks under each condition, participants completed a questionnaire including Likert items about their ability to verify, intervene and steer the AI, sense of control, information overload, frustration levels, and the utility of specific features (exact questions included in Figure \ref{fig:likert_scale_results}). Additionally, participants discussed their experience with each system in a 5-minute semi-structured interview.

The sessions were conducted in-person, lasting approximately 2.5 hours with a short break after using each system. Consent was obtained before running the study and each participant was compensated with a GBP \textsterling50 Amazon gift card. Our study protocol was reviewed and approved by our institution's ethics and compliance review board.

\subsection{Data Collection and Analysis}
We recorded the audio and screen activity during each session using MS Teams. Audio recordings were transcribed for analysis. User interactions and feature usage was also logged.

The think-aloud data was our main source of understanding how participants used the different systems and what they thought about them in comparison with each other. We transcribed the think-aloud data and post-condition interviews, and two researchers performed a negotiated, directed qualitative analysis. Ahead of the analysis, we identified a set of research themes concerning steering and verification during the AI-assisted data analysis process, and report our findings organised by these themes in Section~\ref{sec:results}. Because we were interested in specific themes \emph{a priori}, and were not developing a reusable coding scheme, our analysis differs from the commonly applied inductive approach \cite{braun2006using}. We did not develop a codebook, and this is not a situation in which it is appropriate to seek inter-rater reliability. \highlight{Instead, we used a \textit{``deductive''} coding approach focusing on steering and verification as the main themes.} The two researchers iteratively discussed their interpretation of the findings and negotiated each disagreement until it was resolved \cite{mcdonald2019reliability}.

Task completion was defined as achieving a solution that correctly resolved both issues indicated in Table \ref{tab:final_study_tasks} for each task within the 15-minute time frame. For tasks that were correctly completed, we recorded the number of hints provided during each task and calculated approximate time on task. Task completion time was an approximate of when participants started the task (clicking on the run query button) until they notified the experimenter that they finished the task, with no remaining issues. However, our analysis of task time is only indicative, as think-aloud protocols interfere with accurate timing.

We analyzed post-condition Likert responses to compare the three systems and determine any statistically significant differences using a Friedman Chi Square test on the responses for each question with the system type as the independent variable. When significant differences were found ($\alpha=0.05$), a Wilcoxon signed-rank test identified pairwise significant comparisons. \highlight{Since we made three comparisons (between each pair of systems), w}e applied a Bonferroni correction ($\alpha=0.016$).

\section{Results}
\label{sec:results}
In this section, we present a comparative analysis of the \incremental and \structural AI tools versus the \conversational baseline. Our findings are derived from study observations, log data, participant (P1--P18) think-aloud data, post-condition surveys, and post-study interviews. In turn, we present the results regarding task completion (Section~\ref{sec:results_task_completion}), steering and control (Section~\ref{sec:results_steering}), and verification (Section~\ref{sec:results_verification}).

\subsection{Task Completion}
\label{sec:results_task_completion}
Successful task completion was determined as solving the task with no remaining issues within 15 minutes. Of the 108 task episodes (18 participants $\times$ 6 tasks), only 7 were not completed successfully. P13 had three non-completed tasks, and P3, P8, P11, and P14 each recorded one non-completed task. The distribution of non-completed tasks per condition was as follows: Baseline: 1, \structural: 2, and \incremental: 4. The incidence of task non-completion is too low to permit statistical comparison.

In 31 instances of the 108 task episodes, participants indicated task completion despite remaining issues, indicating insufficient verification. In such situations, the protocol was for the researcher to identify the remaining issue(s), requiring participants to steer the tool towards fixing the problem. A Friedman Chi Square test revealed no statistically significant differences in number of verification hints required across conditions ($F(2,34)=1.0$, $p=.606$), with 13 hints required for Baseline, 10 for \structural, and 8 for \incremental.

Furthermore, a one-way ANOVA showed no significant differences in task completion time between conditions. The mean completion times across conditions indicated that tasks solved with the Baseline tool were finished slightly faster (M=543s, SD=220s), followed by the \incremental tool (M=588s, SD=329s), whereas tasks finished with the \structural tool were solved slightly slower (M=658s, SD=240s).

\begin{figure}[!ht]
    \centering
    \includegraphics[width=1\linewidth]{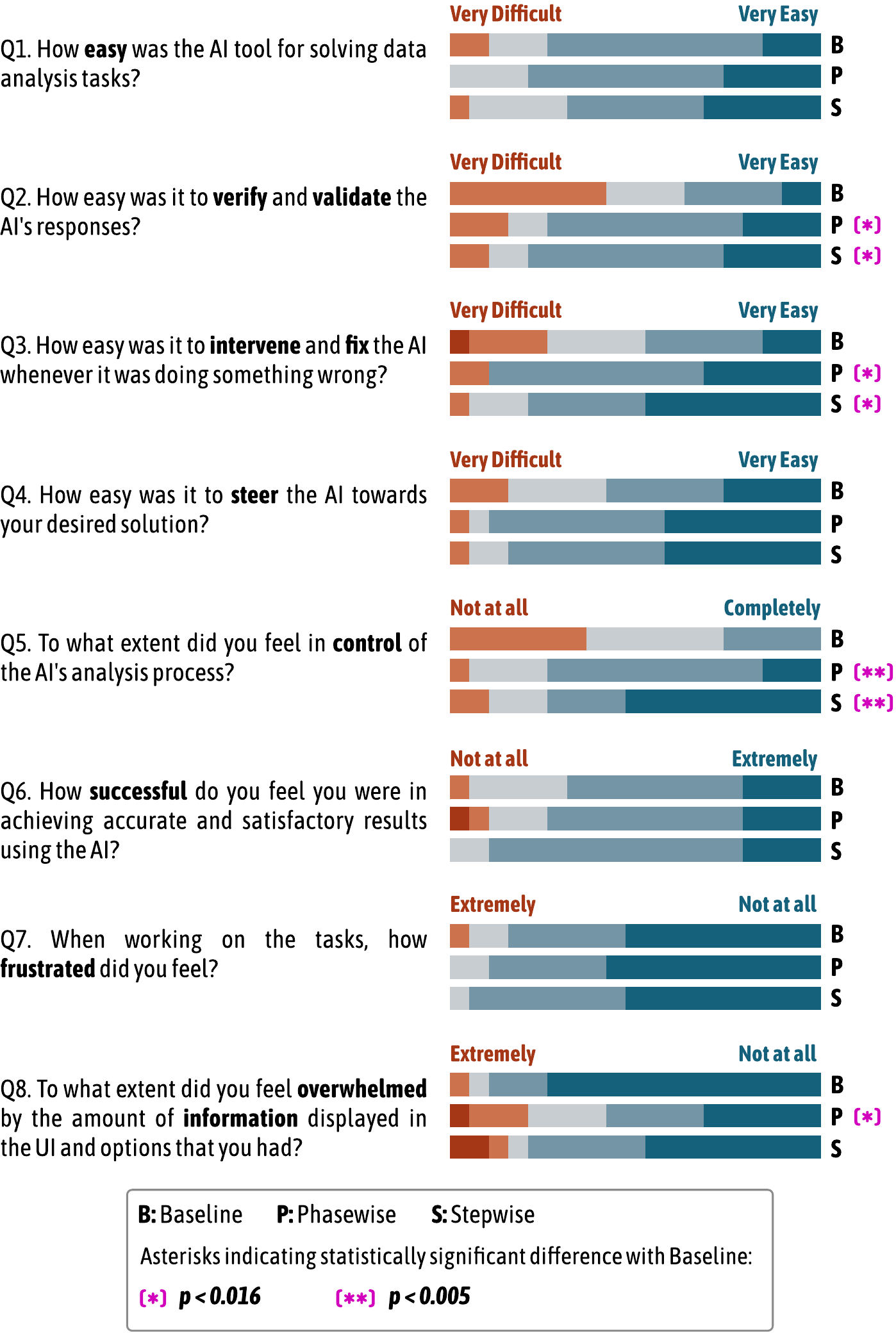}
    \caption{Summary of responses to the post-condition Likert questions for each system.}
    \label{fig:likert_scale_results}
\end{figure}

Finally, post-condition questionnaires on ease of solving EDA tasks (Figure \ref{fig:likert_scale_results}, Q1) or participants' sense of success (Figure \ref{fig:likert_scale_results}, Q6) did not show any statistically significant differences across the three AI tools.

\subsection{Steering and Control}
\label{sec:results_steering}
Analysis of the post-condition questionnaires about control found that participants felt significantly more in control of the AI's analysis process when using the \incremental and \structural systems compared to the Baseline (\incremental-vs-Baseline: $p=.001$, $d=.42$; \structural-vs-Baseline: $p=.004$, $d=.42$). Participants also reported that the \structural and \incremental systems were significantly easier to intervene and fix (Figure~\ref{fig:likert_scale_results}, Q3) whenever it was doing something wrong (\structural-vs-Baseline: $p=.012$, $d=.55$; \incremental-vs-Baseline: $p=.011$, $d=1.05$). However, no significant differences were found in the perceived ease of steering between the three systems (Figure~\ref{fig:likert_scale_results}, Q4).

In the remainder of this section, we explore themes identified within participants' workflows and their think-aloud data. This analysis reveals varied preferences among participants and offers insights into factors that either facilitated or hindered their ability to steer the process. 

\subsubsection{Steering by Directly Editing AI's Assumptions and Actions}
Participants appreciated the ability to directly edit the AI-generated assumptions, thereby aligning the system's operations with their expectations. As P16 stated: 
\textit{``I could add any assumptions that I had in mind and make it the AI's assumptions.''} 
This enabled users to \textit{``steer the AI's decision making process to different directions.''} (P3). For P11, the ability to edit assumptions fostered a more critical perspective towards them; conversely, in the baseline system where the assumptions were fixed, they tended to accept them without questioning, as P11 would \textit{``just go with it as opposed to being critical.''} Furthermore, the baseline's fixed assumptions were a source of frustration, a sense of lack of control, and a barrier to effective interaction (P2, P9, P15, P16, P17). P16 expressed a preference for editing the assumptions directly \textit{``instead of just trying to ask a [follow-up] question,''} and P9 noted that the ability to edit assumptions eliminated the need for manually \textit{``engineering your prompts''}.

The structured editing of assumptions, actions, and execution plan enabled direct manipulation, explicit and fine-grained control over the AI's behavior. P5 expressed that it was easier to interact with, since the assumptions were given and they just had to modify which made the interaction \textit{``less talking and more clicking on buttons''}. Participants appreciated being able to \textit{``edit something very specific''} (P8), such as a step in the execution plan, or \textit{``including the pre-processing steps right beside the columns''} (P7). Similarly, P12 found it difficult to make targeted edits in the absence of structure, stating that \textit{``making small edits [in the Baseline] requires a lot of tweaking.''} Additionally, P6 reported that increased structure facilitated locating information and served as a memory aid. This contrasts with the difficulties they experienced using the \incremental system which lacked the amount of structure used in the \structural system. With the \incremental system, participants had to \textit{``find the correct step to make an edit''} or\textit{``find which column the assumption refers to''} (P6). P4 further indicated that the overall structure provided in the \structural system \textit{``pushes me towards structured analysis''}, specifically \textit{``in terms of validating assumptions''}.

While most participants were generally positive about direct and fine-grained editing of assumptions, P2 and P4 preferred the AI to update its assumptions via natural language queries. P4 additionally criticized the \structural system for the inability to see results update instantly after editing input/output assumptions.

\subsubsection{\highlight{Higher Perceived} Control Through Step-by-Step Task Decomposition}
A distinct advantage observed with the \incremental system was the enhanced control participants reported over the data analysis process. This \highlight{perception} was mainly attributed to tackling the task in smaller, manageable segments. For instance, P16 reported an increased sense of control, by being able to \textit{``easily edit the assumptions and actions in each step.''} Similarly, P11 felt \textit{``much happier''} and \textit{``more confident''}, attributing it to the ability to \textit{``manipulate steps naturally.''} 
P10 also shared a sense of more control over the AI, stating that they \textit{``were not scared''} to make edits to what the AI was doing, as \textit{``it was just a couple of lines,''} and \textit{``it was more inviting to edit the assumptions.''}  
P17 mentioned that the \incremental system facilitated an iterative process \textit{``where [they] could easily go back and change something''.} \highlight{While these reports indicate a perception of enhanced control, actual control should be measured in future work.}

\subsubsection{Steering by Manually Editing Code}
Most participants appreciated the ability to manually edit AI-generated code. P18 mentioned editing the AI-generated code was like \textit{``you're taking over from the AI.''} These edits ranged from minor modifications, such as manually changing a threshold or printing values, to more involved changes such as using the \textit{Generate code} feature to update the logic behind a line of code. When using the \textit{Generate code} feature, participants (P1, P3, P4, P7, P8, P9 P11, P12, P17, P18) selected a line of code and prompted the AI to update it based on a provided natural language query. For instance, P9 selected \inlinecode{df[df['company_location'] == 'US']} in their code and prompted the AI with \texttt{``can you change this line to look for containing `US' instead of strict equality?''} P12 experienced increased control when using the \textit{Generate code} feature since they \textit{``could make very granular prompts''}.

Participants preferred manually editing code for minor changes over using the AI-steering methods provided in each system. Many participants expressed frustration with the inability to directly edit code in the baseline system. P7 stated, \textit{``if [the code] was editable, I can just correct things which I know myself instead of prompting again.''} Notably, P10, unable to edit the AI-generated code directly with the baseline, resorted to copying the desired code line, editing it in the follow-up prompt, and then asking the AI to incorporate the edited code. They found crafting a prompt for the specific edit challenging, admitting, \textit{``I don't really know how to prompt it to get it do what I want.''} However, in some cases, P3 and P10 indicated reasons for \textit{not} wanting to edit the AI-generated code and instead preferred using the AI-steering methods to make the edits. P3 wanted \textit{``to make sure that [their edit] is consistent with the rest of the code''} and P10 stated, \textit{``I don't like editing the code directly when I haven't written it.''}

\subsubsection{Preference for Conversational Steering}
In some instances, participants (P4, P5, P6, P8, and P11) specifically indicated a preference for steering the AI through natural language prompts. They favored the \conversational method of steering the AI over editing the structured assumptions, actions, or execution plans. For example, P4 expressed difficulty in understanding how changes to the assumptions in the \structural system, affected the LLM's output. In contrast, P4 had a more accurate mental model of how to interact with the \conversational baseline, stating \textit{``I know exactly how writing a prompt is going to affect it.''} Others felt constrained by the need to adhere to a specific structure, expressing a preference for more free-form interactions. For instance, P8 described the \conversational system as easier and faster for \textit{``directing the AI using natural language''}, compared to the \incremental system where they felt they were \textit{``trying to change the syntax of the AI.''} Similarly, P5 wanted to \textit{``intentionally write vague prompts and see how much [the AI] understands.''} P4 believed that the \conversational system required less cognitive effort, stating \textit{``I don't like spending that effort to think about it.''} P11 mentioned feeling \textit{``less critical''} about themselves when using the \conversational tool, allowing the AI to \textit{``go and figure it out''} on their behalf. P11 elaborated: \textit{``I knew what it was [that] I wanted it to consider, but when [the tool] is expecting a structured input then I was more concerned with providing it in a nice and structured manner.''}

\subsubsection{Avoiding Edits that Lead to Inconsistency or Regeneration}
To discover participant reasoning behind the selection of a particular steering method from the available options, participants were asked about their specific interactions after each task. We found that participants avoided certain edits when using the \incremental and \structural systems in two cases: 
\begin{itemize}
    \item \highlight{if they had previously edited downstream components and then decided to update an upstream component, the system ignored all downstream edits since regenerations proceed from top to bottom.}
    \item \highlight{if they decided to make edits to downstream components that conflicted with assumptions or code in upstream components.}
\end{itemize}

For example, P1 was worried whether the AI would \textit{``regenerate everything else correctly''} after updating a specific assumption. P12 mentioned that there was \textit{``no obvious way of going back without redoing all of the earlier changes''}. Similarly, P4 mentioned that they \textit{``want to make sure that [their] changes propagate and stay.''} P8 unexpectedly found that they lost downstream edits after making an edit on the upstream components, expressing: \textit{``Oh! So when it regenerated this, it forgot about [their previous edit].''} Participants expressed the need for bidirectional updates to maintain consistency across different components after making an edit. For example, when P18 was using the \incremental system, they could not make edits at the beginning of the problem, which felt natural to them, because \textit{``everything underneath it will drop.''} They preferred that the system would just highlight parts that would be invalidated in the downstream instead of regenerating everything. P10 expressed concern about requiring to \textit{``read everything every time [they] made a small change.''} 
In contrast, P18 accepted previous components going out of sync, as at that point they have \textit{``taken over from the AI''} and P5 appreciated the propagation of changes between components, stating that they liked \textit{``how interconnected things were''}.

\subsection{Verification}
\label{sec:results_verification}
In all systems, participants relied on reading and analyzing the AI-generated code and inspecting the intermediary variables for verification. The \incremental system's approach of breaking down tasks into smaller steps, along with the side conversation feature available in both \incremental and \structural systems, improved participants' confidence in verification. The post-condition questionnaire items indicated that both \incremental  and \structural systems significantly facilitated easier verification (Figure~\ref{fig:likert_scale_results}, Q2) of the generated solution compared to the baseline (\structural-vs-Baseline: $p=.016$, $d=.47$; \incremental-vs-Baseline: $p=.016$, $d=1.44$).

\subsubsection{Verification through Reading Code and Asking Questions}
Reading the code line-by-line was a common verification method. P12 mentioned that \textit{``you still have to read all the code and understand what it's doing''} for verification. 
Participants mostly relied on their own knowledge about Python and Pandas for verification, as stated by P8: \textit{``you have to know how to code, because you have to read the code and make sure it makes sense.''} When P3 was asked how they knew that they had successfully finished the task, they responded \textit{``I inspected the code and found that it handled that edge case correctly.''}
However, in many instances participants had difficulty understanding the AI-generated code, if it used idioms or functions unfamiliar to the participant. 
Participants appreciated the \textit{Ask Question} feature in these situations. 
A majority of participants (n=13) used this feature at least once to explain a portion of the generated code. For example, when P9 was working on Task 2, they stated: \textit{``I've never seen this function \inlinecode{explode()} so I'm just gonna ask what does this do'''}. 
Participants found the responses to their queries useful. For example, P2 asked about the \inlinecode{fillna()} function and realized that the code was doing something undesirable (replacing \inlinecode{nan} with \inlinecode{"Unknown"}). Furthermore, participants felt the absence of the Ask Question feature when using the baseline tool, where for instance P18 wanted to use a search engine, and P6 mixed the main thread of the task with a comprehension question: \textit{``I don't understand [refers to code].''} During the study, several unanticipated, yet effective use cases of the Ask Question feature emerged, such as P4 asking questions about an assumption, and P12 requesting help with debugging by selecting part of the code and asking \textit{``why this code produces an error.''}

\subsubsection{Inspecting Intermediary Variables}
All participants used the intermediary variable inspection feature available in all three systems for verification. Participants inspected variables to \textit{``compare between turns''} (P1), and to \textit{``see if [the system] has done the [operation] correctly''} (P14). P12 mentioned that the verification process was similar to \textit{``debugging''} and P11 indicated that inspecting all the dataframes significantly increases confidence in the process.

\subsubsection{Steering for Verification}
However, generated code was not always easily verifiable. In some instances, the generated code overwrote variables instead of creating new dataframes, which interfered with variable inspection as only the final state of the variable after code execution was displayed. In other cases, the generated code directly computed the final result, without sufficient decomposition of steps necessary for proper verification. For example, P4 mentioned that \textit{``the way that [the system] is generating code does not create useful intermediary dataframes ... it's showing me the end result''}. In some cases this lead to unjustified reliance on the generated code, as P3 mentioned: \textit{``I guess I would need to trust in this case.''} 

Therefore, a recurring theme that emerged was participants trying to update the code, through steering, to include more informative and useful intermediary variables. For example, P16 added a new step to the execution plan to emit new outputs and other relevant columns in addition to just showing the final result. P4 added an explicit step to the execution plan \inlinecode{display couple of groups so I can manually verify}. Interestingly, there were also several cases that participants just did not understand the method used in the generated code, and therefore, asked the system to \textit{``come up with a more understandable solution''} (P6).

\subsubsection{Focusing on Smaller Steps Facilitated Verification}
The \incremental system provided a one-to-one mapping of code with the intermediary variables for each step. Participants found that they can easily \textit{``focus on each small step''} (P15), improved their confidence since they were forced to \textit{``think of edge cases along the way''} (P9). P5 mentioned that \textit{``having it step-by-step leads to more reflecting from my side and verifying each block''}.  Granular decomposition also helped with locating issues. It was \textit{``easier to figure out what is going wrong''} (P11), and \textit{``there was less margin of error''} (P7). For P3, the higher number of intervention points in the \incremental system helped their \textit{``results to be correct all the time.''} 

The step-by-step process was \textit{``more natural''} (P11) and \textit{``felt a lot more similar to how [they] would approach the analysis''} (P9), because \textit{``don't usually solve the whole task at once.''} (P10). Interestingly, P7 \textit{``felt less need for validation since [they are] inspecting after each step''}. 

Additionally, participants experienced less information overload: \textit{``you get the blocks one-by-one so you are not overwhelmed by too much information''} (P5), and compared to the \structural system which felt more like \textit{``debugging somebody else's code than writing my own code''} (P10). However, several participants (P4, P5, P13, and P18) were critical \highlight{of the \incremental system for not providing any information about the upcoming next step.} P18 stated that \textit{``I do not want it to do everything at once, but I want to know what it's gonna do next''} and P5 expressed reduced confidence for \textit{``not knowing the steps in advance.''}


\subsubsection{Aggregated Information Helped with Verification}
Many participants (n=7) appreciated how the \structural system aggregated descriptive statistics and assumptions for each column, \highlight{finding that these elements scaffolded the AI's reasoning about the task.} These helped P15 \textit{``understand what are the different possibilities,''} enabled P6 to \textit{``see exactly how each column will be treated''}, and \textit{``forced''} P3 \textit{``to see the data a bit better.''} For example, in Task 3, P3 easily found the \inlinecode{"Unknown"} genre problem in the descriptive statistics, immediately updating the corresponding assumptions to handle it. Furthermore, P9 justified the usefulness of the aggregated information per columns by stating \textit{``it is something a lot of times you would end up asking about anyways,''} and P1 appreciated that it \textit{``gives you a preview of everything together.''} 

However, some participants mentioned that the aggregated statistics were a source of information overload. The post-condition questionnaires also indicated that they felt significantly overwhelmed (Figure~\ref{fig:likert_scale_results}, Q8) by the amount of information displayed when using the \structural system compared to the baseline ($p=.008$, $d=.11$). P16 felt \textit{``frustrated by the amount of things that [they] saw on the screen''} and P8 stated that \textit{``It could start getting quite cumbersome if the dataset was large''}. Similarly, P4 found the structure to be overwhelming and some assumptions about columns were irrelevant to what they were trying to do, and instead, they wanted the system to contextually display the right amount of information.

\subsubsection{Running Side Queries}
The \textit{Run Side Query} feature of the \incremental and \structural systems was the most frequently used side conversation feature with 82 usages, compared to 26 usages of \textit{Ask Question}, and 17 usages of \textit{Generate Code}. All participants ran side queries at least once. About 75\% (n=62) of the side queries were to understand data and its limitations and 17\% (n=14) were to visualize data for inspection.

The \textit{Side Query} feature facilitated a novel and effective workflow, especially when integrated with the editable assumptions, actions, and execution plan in both systems. Participants used it to \textit{``explore the dataset, validate assumptions, and add them to the column breakdown''} (P17), and \textit{``to build up assumptions and edit the plan''} (P10). P9 used the Side Query to plot the distribution of a column and select a better threshold for filtering outliers. P13 mentioned gaining \textit{``more confidence after plotting histograms''} and found that the Side Query proved more beneficial in the \incremental system because it \textit{``forces me to check the outputs at every step''} in contrast to the \structural system. In the baseline system, without the \textit{Side Query} feature, P4 and P10 resorted to the main thread for their verification queries. This experience led them to appreciate the convenience of having the \textit{Side Query} feature in a side panel, which prevented interference with the main thread. Reflecting on this, P10 highlighted that \textit{``it was not taking away from the history of questions I'd established already''} and expressed a desire to avoid getting \textit{``off track with intervening stuff''} in the baseline system.

\subsubsection{Deferring Steering after Seeing Initial Results}
Participants frequently preferred to first see results before interacting with and steering the AI tool. P6 highlighted that they \textit{``just want to see the result first and then trim it''}. P18 wanted to \textit{``look at the result first and if the result was nonsense then go back''.} This was particularly the case in the \structural and \incremental systems, as P4 mentioned that the system made them \textit{``go through all of these steps and spend so much time before [they] could actually see the result.''} P8 wished to see the generated code because they did not trust the system to generate correct code even if the execution plan was correct: \textit{``part of me wants to just see what it does, even if the [execution] plan looks reasonable, I know there's gonna probably be errors.''} Another reason why participants wanted to defer steering, particularly in the \structural system, was that the input/output assumptions or the execution plan did not have enough details about how the AI is going to \textit{``handle''} or \textit{``calculate''} something (P10, P14). Similarly, P12 was not sure about something in the plan, so they said \textit{``I'm going to generate first and see what happens.''} Lastly, P4 indicated that having a small working memory was why they preferred the baseline system, where every interaction with the AI resulted in a complete output to verify.

\subsection{Summary of Results}
Our study revealed that while there was no difference in task success, completion time, or number of required verification hints, participants felt significantly more in control of the data analysis process when using the \structural and \incremental systems compared to the \conversational baseline. \highlight{Overall, the results show that both systems were preferred over the \conversational baseline. However, because the \structural system led to significantly higher information overload, the \incremental system emerged as the most balanced and effective of the three.}


The study also highlighted the value of side conversations in the AI-assisted data analysis process. The ability to run side queries facilitated an iterative workflow of exploration, validation, and updating of editable assumptions, particularly in the \incremental system. The Ask Question feature helped participants understand the AI-generated code, while the Generate Code feature allowed them to update the logic behind a line of code. In the absence of side conversations, as in the baseline system, participants mixed their queries with the main thread of the task.

In the \structural system, the organization of assumptions enabled direct and broad control and served as a memory aid. Participants appreciated the aggregated descriptive statistics and assumptions for each column, which helped them understand how each column would be treated. However, the amount of information displayed was also a source of overload for some participants.

The \incremental system provided fine-grained control by breaking down tasks into smaller, manageable segments. This improved verification, as participants could focus on each small step and consider edge cases along the way. However, a limitation of the tool was the inability to see the next step in the process.

Finally, some participants preferred the \conversational system for its simpler and more familiar mental model of how asking follow-up questions would affect the AI's output. They also appreciated the flexibility of free-form interactions and the ability to see a result faster. However, the inability to directly edit the AI-generated code was a source of frustration.

\section{Discussion and Implications for AI-Assisted Data Analysis Tools}
Our designs for the \structural and \incremental systems, and their evaluation against the \conversational tool, have provided us with a deeper understanding of the trade-offs within the design space of AI-assisted data analysis tools. This discussion will explore these trade-offs, their impact on user preferences and interactions, and suggest guidelines for design.

Our study finds that the key to designing AI-assisted data analysis tools lies in providing the user with the necessary controls to make informed decisions and maintain control over the process. This echoes the longstanding positioning of the role of user interface elements in interactive machine learning systems as providing \emph{decision support} \cite{kocielnik2019will,sarkar2015interactive,sarkar2016phd}. Our study finds that in the specific case of AI-assisted data analysis, decision support is subject to the following key design questions:

\begin{itemize}
    \item \textbf{DQ1 Steering Points:} At what points should the system allow the user to intervene in the process and steer? How frequently should these steering opportunities occur?
    \item \textbf{DQ2 Steering Support:} How does the user \highlight{verify the current state of the AI's output} and determine which direction they should steer it? How should the tool facilitate users in making informed decisions at each steering point?
    \item \textbf{DQ3 Steering Modality:} What interface affordances are available for the user to steer the process? How structured or flexible should the modality of their interaction be?
\end{itemize}

These are similar to the design questions regarding the number and nature of ``choice points'' within a data analysis workflow generated by an earlier generation of tools termed Intelligent Discovery Assistants (IDAs) \cite{serban2013survey}. We find that the choices users face with IDAs (e.g., what type of regression or normalisation to apply at a particular step) still exist within generative AI-assisted data analysis, but they are embedded within the higher-level challenges of steering, and are experienced by users as a secondary concern.

\subsection{DQ1: Steering Points}
One design question is at which points during the generation process the user should reflect, check for correctness, and steer if needed. Our \incremental and \conversational tools can be seen as two ends of a spectrum of intervention opportunities. The \incremental system offers steering points after each step of the analysis, allowing for incremental adjustments. In contrast, the \conversational tool aims to complete the task with minimal user interruption, offering a chance to adjust only after attempting to solve the task.

Our results comparing user experiences with these systems reveals significant trade-offs. More frequent steering points increase users' confidence in the results and their sense of control over the AI's process. However, it demands more cognitive effort and delays the final outcome due to the frequent pauses created by the intervention points. Additionally, it may lead to premature decisions as users commit to directions without understanding their future impact. According to the Cognitive Dimensions of Notations framework \cite{green1989cognitive}, this is a clear case of the system imposing ``premature commitment''. Indeed, our findings indicate varied user preferences between the \conversational and \incremental systems due to these factors.

To address this challenge and balance control with cognitive load, AI-assisted data analysis tools could adopt a strategy that initially attempts to solve the task without user intervention, followed by a more interactive steering and verification phase with a rich set of steering points. This is similar to the highly validated design strategy of information visualization tools to provide an ``overview first'', and only later ``details on demand'' \cite{shneiderman2003eyes}. 


%

\subsection{DQ2: Steering Support}
At each steering point, users must verify whether the output is correct, and if not, choose a steering action. How should the tool facilitate the information seeking and exploration process that is required for the user to make informed decisions? Tools for verifying AI-generated content are termed ``co-audit'' \cite{gordon2023co}. Auditing and verification are part of the analyst process of sensemaking \cite{pirolli2005sensemaking} and information foraging \cite{pirolli1999information}. Critically, such processes are an opportunistic mix of top-down hypothesis formation and bottom-up hypothesis testing \cite{drosos2024duck}, with multiple activities proceeding in a parallel, non-linear fashion. This is antagonised by the sequential nature of chat interfaces.

Our exploration into the design space introduced side conversations and the \textit{Run Side Query} function to aid this process. Furthermore, in the \structural system, the system displayed dataset columns relevant to the user's task, along with interactive descriptive statistics. Moreover, the execution plan component suggested optional steps for the user to consider adding before proceeding to the next step, enhancing the decision-making process.

Our findings indicate that when the system provides timely, accurate, and relevant information, it fosters a genuinely collaborative experience. Conversely, displaying irrelevant information can reduce trust in the AI and potentially leading to information overload. Users also risk becoming overly dependent on AI for guidance, potentially neglecting critical information seeking which may lead to poor decision making.

Participants noted their desire for the AI to act as an agent, aiding in the assumption-building process at steering points. Future tools could automatically retrieve assumptions by pinpointing specific, relevant evidence to support informed decision-making without overwhelming users. Additionally, these tools could help in accessing domain-specific knowledge pertinent to the data analysis task. Lastly, tools should be transparent regarding the AI's limitations in sourcing all necessary information for optimal decision-making at each decision point.

\subsection{DQ3: Steering Modality}
After the user has decided which direction to steer the AI, their next action is to specify their intent to the AI. The \highlight{design} question here is determining the right interface and modality for the user to steer the AI. In response, we introduced two distinct interfaces: a free-form text editor used in the \conversational system for maximal flexibility, and structured editors in the \incremental and \structural systems. The structured editors contain the assumptions and actions generated by the LLM's chain-of-thought, allowing users to edit them for steering.

Our results highlight the trade-offs between these approaches. An unstructured and flexible modality reduces perceived cognitive load, and allows users to be less self-critical when making edits. Users can form a more straightforward mental model of how their inputs steer the AI and receive immediate feedback. However, increased flexibility shifts the responsibility of precise and effective interaction onto the user. Users lose fine-grained control and may have to engage in the challenging and time-consuming task of prompt design \cite{liu2023wants, zamfirescu2023johnny}.

Depending on the generative AI and data analysis expertise of the user, a range of steering methods may be appropriate. A potential approach could be adding the ability to switch between structured editing of assumptions or instructing the AI with free-form queries. Moreover, to increase transparency in the user's mental model about how their edits affect the system, tools could enable inspection of the underlying LLM prompts, and highlight how their steering edits affect the information sent to the model. Lastly, advanced users might appreciate the ability to manually adjust the underlying prompts. These design suggestions are complementary to established prompting guidelines and practices (e.g., \cite{mishra2021reframing}).

\section{Limitations}
We identified several limitations in our study and system design that should be considered when interpreting the results.

\subsection{Study Limitations}
In our evaluation study, the tasks were manually made more complex and less clean to always include errors when presented to the AI tools to require further verification and steering. However, this may have impacted the ecological validity of the the tasks. Another challenge to ecological validity is our decision to provide the initial NL query for each task, which does not reflect how these systems would be used in practice, and reduced the opportunity for us to study the consequences of divergent natural prompting strategies. For our study, this was an acceptable trade-off as it guaranteed that all participants would encounter the same steering and verification needs, which allowed clearer comparisons between the different systems. It also allowed us to sidestep the issue of participant queries being unevenly ``primed'' by the task description \cite{liu2023wants}. Future work may relax this constraint to study a wider range of participant prompting strategies.

Typicality and novelty preferences may have influenced how participants ranked their preference of features or system \cite{hekkert2003most}. Participants might also be biased towards systems they believe are of personal interest to the researcher (known as the ``yours is better'' bias) \cite{dell2012yours}. As a mitigating measure, the researcher did not associate themselves with the prototypes in this study and elicited reflections grounded in participants' concrete experiences rather than subjective perceptions \cite{bernhart1999patient}. These reflections were further corroborated through screen recordings and usage logs \cite{czerwinski2001subjective,mackay1998triangulation}.

Participants only used each system for a short time (30 minutes per system, not including the tutorial), which is typical of controlled experiments in laboratory settings. These cannot capture long-term effects \cite{sarkar2023simplicity}; some phenomena only emerge over long-term use and some phenomena which appear to be salient with short-term use erode over time. Consequently, future work could aim to cross-validate our findings longitudinally using experience sampling \cite{csikszentmihalyi1987validity} or diary studies \cite{rieman1993diary}.

\subsection{System Limitations}
During the study, we observed limitations in how the system propagated users edits, both upstream and downstream. The first limitation involved propagating changes from edited assumptions to generated code, where in some cases the language model would appear to ignore the edits. This is a fundamental failure mode of generative AI systems, however, system implementations and interfaces can exacerbate the issue. Complex prompting approaches with many instructions can make it unlikely that the model identifies small changes to assumptions. Additionally, distinguishing assumptions in the interface can set incorrect user expectations around how a model attends to assumptions.

Another limitation of the system is that edits to downstream code or assumptions are not propagated to upstream assumptions, and if a user makes an earlier edit it will overwrite any subsequent changes. The intention behind this \emph{prima facie} design decision was to present a simple model of ``cause and effect'' that represented how completions were generated, namely, the \emph{context} of any point in the system is only that which appears before. Some participants identified this limitation and it influenced their steering preferences, choosing not to edit an assumption because it would clear changes that had been made to code.

\section{Conclusion}
In this work, we explore the design space of AI-assisted data analysis tools by presenting two novel interfaces that aim to improve steering and verification. Starting from the observation that task decomposition is an emerging characteristic of recent LLM-based systems, we developed two systems that explore different modes of interactive task decomposition, each based on unique trade-offs. The first, \incremental, decomposes the problem step by step; the second, \structural, decomposes the problem \highlight{into logical} phases.
Our evaluation demonstrates that users experienced a greater sense of control and confidence with our systems in comparison to a chat-based baseline. Still, task decomposition is not without preference or cost. Some users prefer to work through the task incrementally, whilst others prefer to see the plan upfront. Additionally, highly-structured decomposition can introduce cognitive burden.
Consequently, we imagine that future AI interfaces will need to support adaptive decomposition that reacts to the user and task.

\bibliographystyle{ACM-Reference-Format}
\typeout{}
\bibliography{references}

\end{document}